\documentclass[aps,pre,twocolumn,showpacs,superscriptaddress]{revtex4-1}

\usepackage[utf8]{inputenc}
\usepackage[T1]{fontenc}
\usepackage[german,english]{babel}
\usepackage{amsmath}
\usepackage{mathtools}
\usepackage{amssymb}
\usepackage{bbm}
\usepackage{graphicx}
\usepackage{hyperref}
\usepackage[arrow, matrix, curve]{xy}
\usepackage{bm}
\usepackage{yhmath}
\usepackage{color}
\usepackage{url}
\usepackage{mathrsfs}
\usepackage{verbatim}
\usepackage{tabularx}
\usepackage{textcomp}

\usepackage{braket}
\newcommand\diff{\mathrm{d}}

\renewcommand\vec[1]{\boldsymbol{\mathrm{#1}}}
\renewcommand\i{\text{i}}
\newcommand\e{\text{e}}

\usepackage[usenames,dvipsnames]{xcolor}
\hypersetup{colorlinks=true, linkcolor=BrickRed, urlcolor=blue!50!black, citecolor=blue!50!black}

\usepackage[normalem]{ulem}

\makeatletter
\newcommand\hide@visible[1]{%
  \bgroup\fboxsep=.3ex\colorbox{Gray}{begin hide}%
  #1\colorbox{Gray}{end hide}\egroup%
}
\newcommand\hide@hidden[1]{%
  \bgroup\fboxsep=.3ex\colorbox{Gray}{hidden text}%
}
\newcommand\hide@invisible[1]{}
\newcommand\makevisible{\let\hide\hide@visible}
\newcommand\makehidden{\let\hide\hide@hidden}
\newcommand\makeinvisible{\let\hide\hide@invisible}
\makeatother
\makehidden

\begin{document}


\title{Two-dimensional Brownian motion of anisotropic  dimers }




\author{Daniel B. Mayer} 
\affiliation{Institut f{\"u}r Theoretische Physik,  Universit\"at Innsbruck,
Technikerstra{\ss}e 25/2, A-6020 Innsbruck, Austria}

\author{Erick Sarmiento-G{\'o}mez}
\affiliation{Condensed Matter Physics Laboratory, Heinrich Heine University, Universit\"atsstra{\ss}e 1, 
D-40225 D\"usseldorf, Germany}
\affiliation{Divisi{\'o}n de Ciencias e Ingenierias, Departamento de Ingenieria F\'{i}sica, Universidad de Guanajuato, Le{\'o}n, Mexico}

\author{Manuel A. Escobedo-S{\'a}nchez}
\affiliation{Condensed Matter Physics Laboratory, Heinrich Heine University, Universit\"atsstra{\ss}e 1, 
D-40225 D\"usseldorf, Germany}

\author{Juan Pablo Segovia-Guti{\'e}rrez }
\affiliation{Condensed Matter Physics Laboratory, Heinrich Heine University, Universit\"atsstra{\ss}e 1, 
D-40225 D\"usseldorf, Germany}

\author{Christina Kurzthaler} 
\affiliation{Department of Mechanical and Aerospace Engineering, Princeton University, Princeton, NJ 08544, USA}

\author{Stefan U. Egelhaaf}
\affiliation{Condensed Matter Physics Laboratory, Heinrich Heine University, Universit\"atsstra{\ss}e 1, 
D-40225 D\"usseldorf, Germany}

\author{Thomas Franosch} 
\affiliation{Institut f{\"u}r Theoretische Physik,  Universit\"at Innsbruck,
Technikerstra{\ss}e 25/2, A-6020 Innsbruck, Austria}

\date{\today}

\begin{abstract}
We study the 2D motion of colloidal dimers by single-particle tracking and compare the experimental observations obtained by bright-field microscopy to theoretical predictions for anisotropic diffusion. The comparison is based on the mean-square displacements in the laboratory and particle frame as well as  generalizations of the self-intermediate scattering functions, which provide insights into  the rotational dynamics of the dimer. The diffusional anisotropy leads to a measurable translational-rotational coupling that becomes most prominent by 
aligning the coordinate system with the initial orientation of the particles. In particular, we find a splitting of the time-dependent diffusion coefficients parallel and perpendicular to the long axis of the dimer which decays over the orientational relaxation time.
Deviations of the self-intermediate scattering functions from pure exponential relaxation are small but can be resolved experimentally. The theoretical predictions and experimental results agree quantitatively. 
\end{abstract}

\maketitle

\section{Introduction}

The erratic motion of a micrometer-sized particle suspended in a fluid, usually referred to as Brownian motion, has been rationalized by Einstein in his \emph{annus mirabilis} (1905) in terms of incessant collisions with the fluid molecules~\cite{Einstein:AnnPhys_4:1905}. Using the newly developed ultramicroscope, Jean-Baptiste  Perrin and his students subsequently provided  precision data in a \emph{tour de force} of  single-particle tracking experiments \cite{Perrin:AnnChemPhys_18:1909} confirming quantitatively Einstein's predictions. These experiments were rewarded with the Nobel Prize in Physics in 1926. 

The translational motion of  a single spherical particle in a fluid can be characterized in terms of a Gaussian propagator with the mean-square displacement increasing linearly in time~\cite{Einstein:AnnPhys_4:1905, Doi:Theory_of_Polymer_Dynamics:1986}. Nevertheless, subtle persistent correlations due to hydrodynamic memory have been reported at time scales where momentum has not yet 
relaxed~\cite{Lukic:PRL_95:2005,Jeney:PRL_100:2008,Franosch:Nature_478:2011,Huang:NatPhys_7:2011}. 

More complex transport phenomena occur for anisotropic particles, which have been  addressed in the early work of 
Francis Perrin~\cite{Perrin:JPRadium_5:1934,Perrin:JPRadium_7:1936}, 
the son of  Jean-Baptiste Perrin. In this case, the diffusion coefficient is replaced by a symmetric matrix which becomes diagonal in the principle-axis frame and the different eigenvalues encode 
the dependence of the translational diffusion on the particle orientation. Due to the additional rotational diffusion the corresponding Langevin equations or the associated Smoluchowski-Perrin equation become non-trivial~\cite{Doi:Theory_of_Polymer_Dynamics:1986}.

The effects of the translational-rotational (TR) coupling  on observables, such as the mean-square displacement (MSD) in the particle frame  or the self-intermediate scattering function (ISF),  are small. 
Nevertheless, the implications of the TR coupling have been directly visualized by Han \emph{et al.}~\cite{Han:Science_314:2006} in single-particle-tracking experiments of ellipsoidal  particles undergoing Brownian motion  in 2D. In particular, they resolved the orientational motion of the particle in addition to the translational motion allowing them to measure correlation functions 
with restrictions on the initial orientation. In terms of these more general correlation functions they were able to  highlight non-trivial aspects of the anisotropic 
diffusion~\cite{Han:Science_314:2006,Han:PRE_80:2009, Zheng:JCP_133:2010}. However, their analysis was limited to the low-order moments of the translational and rotational displacements and did not consider spatio-temporal information as encoded in the self-intermediate scattering function~\cite{Doi:Theory_of_Polymer_Dynamics:1986}. 

In addition to the motion of elliptical particles, the dynamics of other anisotropic particles has been studied, such as dimers or tetrahedral clusters \cite{Vivek:JCP_147:2017, Edmond:PNAS_109:2012, Sarmiento:PhysRevE_94:2016}.
Recently, microfabricated `boomerang particles' with unequal arm lengths have been investigated as model systems for asymmetric particles undergoing 2D Brownian motion~\cite{Chakrabarty:Langmuir_30:2014,Chakrabarty:SM_12:2016,Koens:SM_13:2017}.  Non-trivial mean-square displacements have been found if the tracking point does not coincide with the center of hydrodynamic stress (CoH). This becomes also relevant for non-rigid particles, such as elastic macromolecules~\cite{Cichocki:JFM_878:2019}. Moreover, for flexible colloidal trimers additional hydrodynamic couplings between conformational changes and the translational motion have been demonstrated very recently~\cite{Verweij:PRR_2:2020}.

Single-particle tracking based on video and confocal microscopy allows  monitoring 
many particles simultaneously with a high spatial and temporal resolution~\cite{Manzo:RoPP_78:2015,Crocker:JCIS_179:1996, Jenkins_ACIS_136:2008, Prasad:JPCM_19:113102, Vivek:JCP_147:2017, Kurzthaler:PRL_121:2018, Wang:PNAS_106:2009, Chakrabarty:Langmuir_29:2013, Edmond:PNAS_109:2012, Anthony:AnalMethod_7:2020}. It has become a standard tool to directly measure individual trajectories and provide access to the translational and rotational motion of colloidal particles. 
The diffusion anisotropy moreover can be extracted from particle trajectories using advanced statistical methods~\cite{Hanasaki:PRE_85:2012, Matsuda:AnalChem_88:2016}.
Furthermore, 
the translational and rotational motion of colloidal particles that are anisotropic either in shape or optical properties have been quantitatively characterized using depolarized dynamic light scattering (DLS) 
\cite{Berne:Dynamic_Light_Scattering:2000, Pecora:JNR_2:2000, Aragon:JCP_82:1985, Degiorgio:ACIS_48:1994, Degiorgio:PRE_52:1995, Diaz:JCP_121_2004}.
Depolarized DLS has been complemented by near-field depolarized DLS to study translational and rotational dynamics on a larger length scale \cite{Brogioli:OE_17:2009, Escobedo:JCP_17:2015}.
Combining  microscopy and scattering approaches, in movies obtained using a polarization or dark field microscope the temporal correlations of the spatial Fourier modes of the intensity fluctuations can be analyzed. These 
techniques, polarized and dark-field differential dynamic microscopy, also yield
 the translational and rotational dynamics~\cite{Giavazzi:JPCM_28:2016, Cerbino:JPCM_30:2018}. 
Differential dynamic microscopy (DDM) \cite{Cerbino:PRL_100:2008, Sentjabrskaja:NatComm_7:2016, Wilson:PRL_106:2011, Giavazzi:JOpt_16:2014, Giavazzi:PRE_80:2009}
has also been applied to determine the anisotropic diffusion of anisotropic particles in an 
external field \cite{Reufer:Langmuir_28:2012, Pal:SciAdv_6:2020}.

Here, we follow  individual dimers performing Brownian motion in a quasi-2D geometry and compare the experimental observations with analytical predictions for anisotropic diffusion.  Single-particle tracking  allows us to resolve the orientation of the dimer and thus  to extract directly the diffusional anisotropy as well as the rotational self-diffusion coefficient. We further determine the transient
dynamic splitting of the anisotropic diffusion for a restricted initial angle following 
Ref.~\cite{Han:Science_314:2006}. We extend the analysis by deriving theoretical predictions for the generalized self-intermediate scattering functions which  renders the dynamics on different length scales accessible. This spatio-temporal information provides more information on the dynamics than the low-order moments and yields further insights into the TR coupling. 

The manuscript is organized as follows: 
In Sec.~\ref{Sec:Theory} we introduce the theoretical model  and elaborate the predictions for 
relevant observables while Sec.~\ref{sec:MM} provides the experimental details. Readers mainly interested in the results may jump directly to Sec.~\ref{Sec:Results}, where  
the experimental results are presented and compared to the  theoretical predictions. Last, we give final remarks in Sec.~\ref{Sec:Summary}.

\section{Theory}\label{Sec:Theory}

In  this section we  introduce the model of planar anisotropic Brownian motion and  review theoretical results for measurable quantities. In particular, we elaborate on the consequences of the translational-rotational (TR) coupling. Many of the results can be  found in the literature, for example, the lowest non-trivial moments 
of the dynamics~\citep{Han:Science_314:2006}, the formal solution of the associated Fokker-Planck equation in terms of Mathieu functions for the conventional self-intermediate scattering function~\cite{Munk:EPL_85:2009}, and expressions for the generalized self-intermediate scattering function~\cite{Munk:PhD_thesis:2008}. However, considerations on the consequences of symmetries and equilibrium dynamics appear not to exist in the literature so far.  

\subsection{Model} 

We consider the dynamics of a passive Brownian particle confined to a plane. Its instantaneous orientation $\vec{U}(t)=(\cos\Theta(t),\sin\Theta(t))^{\text{T}}$, parametrized by the polar angle $\Theta(t)$, undergoes rotational diffusion characterized by the rotational self-diffusion coefficient $D_{\text{rot}}$. Additionally, the motion of the particle is subject to anisotropic translational diffusion encoded via the short-time self-diffusion coefficients parallel $D_{\parallel}$ and perpendicular $D_{\perp}$ to its orientation. In the current planar case we can choose the orientation such that $D_\parallel \geq D_\perp$. 
 The equations of motion (e.o.m.) are  Langevin equations for the fluctuating position $\vec{R}(t)=(X(t),Y(t))^{\text{T}}$ and the fluctuating angle $\Theta(t)$. They  read in It\={o} form~\cite{Doi:Theory_of_Polymer_Dynamics:1986}
\begin{subequations}
\begin{align}
\frac{\diff\Theta(t)}{\diff t}&=\sqrt{2D_{\text{rot}}}\psi(t), \label{eq:LangevinAngle} \\
\frac{\diff\mathbf{R}(t)}{\diff t}&=\sqrt{2D_{\perp}}\boldsymbol{\xi}(t) \nonumber \\&+\left(\sqrt{2D_{\parallel}}-\sqrt{2D_{\perp}}\right) 
\vec{U}(t)\left(\vec{U}(t)\cdot\boldsymbol{\xi}(t)\right),\label{eq:LangevinPos}
\end{align}
\end{subequations}
where the random fluctuations are modeled in terms of  independent Gaussian white-noise processes $\psi(t)$ and $\boldsymbol{\xi}(t)=(\xi_x(t),\xi_y(t))^{\text{T}}$ with zero mean and delta-correlated variances $\langle\psi(t)\psi(t^{\prime})\rangle=\delta(t{-}t^{\prime})$ and $\langle\xi_{i}(t)\xi_{j}(t^{\prime})\rangle=\delta_{ij}\delta(t{-}t^{\prime})$. 

For anisotropic diffusion, $ D_\parallel > D_\perp$, Eq.~\eqref{eq:LangevinPos} encodes a non-trivial coupling of the position $\vec{R}(t)$ to the orientation $\vec{U}(t)$ such that the pure translational motion no longer corresponds to simple diffusion. The main purpose of this section is to elaborate the consequences of the TR coupling on measurable quantities. 

The model contains three transport coefficients, $D_{\text{rot}}, D_\parallel, D_\perp$, which we treat  as phenomenological parameters. The rotational diffusion time $\tau_{\text{rot}} := 1/D_{\text{rot}}$ sets the characteristic time scale for the spatio-temporal dynamics and can be inferred from Eq.~\eqref{eq:LangevinAngle}. Essentially at this time scale the TR coupling relaxes and a crossover to isotropic diffusion occurs with the mean translational self-diffusion coefficient $\bar{D} := (D_\parallel+ D_\perp)/2$  as long-time diffusion coefficient.
An intrinsic length scale  can be extracted by $\ell:=\sqrt{ 3 \bar{D} /4 D_{\text{rot}}}$, where the prefactor is chosen such that 
$\ell$ corresponds to the geometric radius of a spherical particle immersed in a 3D bulk viscous fluid with diffusion coefficients $D_{\text{rot}} = k_B T/8 \pi \eta \ell^3$ and $ \bar{D} = k_B T/6\pi \eta \ell$. In total, two of the parameters set the scale for length and time, while the anisotropy $\Delta D / \bar{D} \in [0,1]$ as ratio between the translational anisotropy $\Delta D:=(D_\parallel- D_\perp)$ and the mean translational self-diffusion 
coefficient $\bar{D}$, is the only dimensionless parameter of the problem.

For a general reference point on the rigid particle, additional couplings between the rotational noise and the translational motion as well as the translational noise and the rotational motion occur. However in 2D one can always find a reference point called center of hydrodynamic stress (CoH), such that these couplings evaluate to zero. Our e.o.m., Eqs.~\eqref{eq:LangevinAngle} and \eqref{eq:LangevinPos}, therefore refer to the CoH. 
Correspondingly, we will check that the tracking point agrees with the CoH to avoid correlations~\cite{Chakrabarty:Langmuir_30:2014,Chakrabarty:SM_12:2016,Koens:SM_13:2017} that complicate the analysis (Sec.~\ref{CoM-CoH}).

Let us summarize some symmetries of the dynamics. The e.o.m.\@ reflect isotropy in a statistical sense, i.e.~for each trajectory in 
configuration space $(\vec{R}(t), \Theta(t))_{t\in \mathbb{R}}^{\text{T}}$ there is an equally likely rotated trajectory $( \textsf{R}_\chi \cdot \vec{R}(t) , \Theta(t)+ \chi)_{t\in \mathbb{R}}^{\text{T}}$ where ${\sf R}_\chi$ is the 2D rotation matrix rotating counter-clockwise by an angle $\chi$. 
Next, the dynamics is non-chiral meaning  that reflecting a trajectory, e.g.~at  the $x$-axis $(X(t),Y(t), \Theta(t))_{t\in\mathbb{R}}^{\text{T}}\mapsto (X(t),-Y(t), - \Theta(t))_{t\in\mathbb{R}}^{\text{T}}$, yields a new trajectory with  the same probability. 
Moreover, since Eq.~\eqref{eq:LangevinPos} contains only even powers of the orientation $\vec{U}(t)$, a rotation of the orientation of the particle by $\pi$ leads to identical trajectories for the same noise  history.
Furthermore the e.o.m.\@ encode an equilibrium dynamics, which means that the time reversed trajectory enters with equal weight in the ensemble. 
The last two properties are  violated for example by an active Brownian particle
~\cite{Kurzthaler:SciRep_6:2016, Kurzthaler:PRL_121:2018, Volpe:AJP_82:2014, tenHagen:JoP_23:2011} and an active circle swimmer would additionally display chiral dynamics~\cite{Teeffelen:PRE_78:2008, Kurzthaler:SM_13:2017, Kuemmel:PRL_110:2013}. 

\subsection{Relevant observables}
The simplest quantities involve only statistical properties of the fluctuating displacement $\Delta \vec{R}(t) := \vec{R}(t)-\vec{R}(0)$, for example, the mean-square displacement $\langle |\Delta \vec{R}(t)|^2 \rangle$ or the mean quartic displacement $\langle |\Delta \vec{R}(t)|^4 \rangle$.
Here the brackets $\langle \ldots \rangle$ indicate an average over the noise or equivalently a time 
and/or ensemble average along/over the trajectories. More information on the spatio-temporal behavior of the translational motion is encoded in the characteristic function of the random displacements, i.e. the self-intermediate scattering function (ISF)
\begin{align}\label{eq:convISF}
F(k,t) := \langle \exp(- \i\, \vec{k} \cdot \Delta \vec{R}(t) ) \rangle ,
\end{align}
where $\vec{k}$ is referred to as scattering vector. 
By statistical isotropy the ISF is independent of the direction of $\vec{k}$, it only depends  on the wavenumber $k = |\vec{k}|$. 
In dilute suspensions of Brownian particles the ISF can be directly measured, for example in homodyne or heterodyne scattering experiments~\cite{Berne:Dynamic_Light_Scattering:2000, Aragon:JCP_82:1985}, by single-particle tracking~\cite{Kurzthaler:PRL_121:2018, Wang:PNAS_106:2009, Chakrabarty:Langmuir_29:2013},  or by differential dynamic microscopy 
(DDM)~\cite{Cerbino:PRL_100:2008, Sentjabrskaja:NatComm_7:2016, Kurzthaler:PRL_121:2018, Wilson:PRL_106:2011, Giavazzi:JOpt_16:2014, Giavazzi:PRE_80:2009}.

We  take advantage of the possibility to monitor  the angular motion $\Theta(t)$ in single-particle-tracking experiments and define more general observable quantities that are sensitive also to the orientation of the particle. Any  two-time correlation function can be calculated if the propagator 
\begin{align}
\mathbb{P}( \vec{r}, \vartheta, t | \vartheta_0) 
:= \left\langle \delta\left(\vec{r}{-}\Delta \vec{R}(t) \right) \delta_{2\pi}( \vartheta{-}\Theta(t) )  \right\rangle_{\vartheta_0}
\end{align} 
is known, i.e.\@ the conditional probability density that a particle with initial  orientation $\vartheta_0 $ is displaced by $\vec{r}$ in lag-time $t$ and displays a final orientation $\vartheta$.  The angular delta-function $\delta_{2\pi}(\cdot) := \delta( \cdot \text{ mod } 2\pi)$ reduces the  argument modulo $2\pi$. Moreover, conditional averages with respect to the initial angle are abbreviated as $\langle \ldots \rangle_{\vartheta_0} := 2\pi \langle \ldots \delta_{2\pi}(\vartheta_0{-}\Theta(0))  \rangle$,  where we have taken into account that the probability for the initial angle $\vartheta_0$ is uniform on the circle. For zero lag time 
the initial condition  
\begin{equation}
\mathbb{P}(\vec{r},\vartheta,t{=}0 | \vartheta_{0})=\delta(\vec{r})\delta_{2\pi}(\vartheta{-}\vartheta_{0}),
\label{eq:Pini}
\end{equation}
is imposed.
For example, the moments of the displacement can be calculated via
\begin{align}
\langle |\Delta \vec{R}(t)|^{2n} \rangle = \int_0^{2\pi} \!\!\diff \vartheta \int_0^{2\pi}\!\! \frac{\diff \vartheta_0}{2\pi} \int_{\mathbb{R}^{2}} \diff^{2}  r \, |\vec{r}|^{2n} \mathbb{P}(\vec{r}, \vartheta, t | \vartheta_0 )  ,
\end{align} 
 where the integral over $\vartheta_0$ corresponds to averaging over the initial orientation, and the integral over $\vartheta$ to summing over all final orientations.
 
 Generalizations of the self-intermediate scattering functions are then introduced as Fourier transforms with respect to the displacement
\begin{align}\label{eq:Spatial_Fourier}
\hat{\mathbb{P}}(\vec{k}, \vartheta, t| \vartheta_0 ) &:=  \int_{\mathbb{R}^{2}} \diff^{2} r \, \exp(-\i \vec{k} \cdot \vec{r})  \mathbb{P}(\vec{r}, \vartheta, t | \vartheta_0 ) \nonumber \\
&= \langle \exp(-\i \vec{k}  \cdot \Delta \vec{R}(t)) \delta_{2\pi}( \vartheta{-}\Theta(t) ) \rangle_{\vartheta_0} .
\end{align} 
 By statistical isotropy, this generalized ISF is invariant under a simultaneous rotation of the scattering vector $\vec{k}$ and the initial $\vartheta_0$ and final angle $\vartheta$. Setting the lag time to zero yields as initial condition 
\begin{align}
\hat{\mathbb{P}}(\vec{k},\vartheta, t{=}0| \vartheta_0 ) = \delta_{2\pi}(\vartheta{-}\vartheta_0) . 
\label{eq:Pkini}
\end{align}
The dependence on the final angle may be decomposed into Fourier modes  
\begin{align}
\hat{\mathbb{P}}(\vec{k},\vartheta,t| \vartheta_0) = \frac{1}{2\pi} \sum_{\nu=-\infty}^\infty F_\nu(\vec{k},t|\vartheta_0) \e^{\i \nu \vartheta}
\end{align}
with expansion coefficients 
\begin{equation}\label{equ:ISF1}
F_{\nu}(\vec{k},t\vert\vartheta_{0}) : = 
\langle\exp(-\i\left(\vec{k}\cdot\Delta\vec{R}(t)+\nu \Theta(t)\right)\rangle_{\vartheta_{0}}
\end{equation}
for each mode index $\nu\in \mathbb{Z}$. We refer to $F_\nu(\vec{k}, t |\vartheta_0)$ as the restricted ISF.  Its initial values then read $F_\nu(\vec{k},t{=}0| \vartheta_0) = \exp(-\i \nu \vartheta_0)$. 
Similarly, we decompose the dependence on the initial angle in Fourier modes by defining  the generalized ISF 
\begin{align}\label{eq:ISF2}
&F_{\mu\nu}(\vec{k}, t):=\int_{0}^{2\pi}\frac{\mathrm{d}\vartheta_{0}}{2\pi}\e^{\i\mu\vartheta_{0}}F_{\nu}(\mathbf{k},t|\vartheta_{0})  \\
     &\qquad\quad=\langle\exp(-\i\left(\vec{k}\cdot\Delta\vec{R}(t)-\mu\Theta(0)+\nu\Theta(t)\right))\rangle . \nonumber
\end{align}
Its initial values $F_{\mu\nu}(\vec{k},t{=}0) = \delta_{\mu\nu}$ follow by averaging  over the uniformly distributed initial angle $\Theta(0)$.  
Then the propagator is obtained by the double Fourier expansion
\begin{align}
\hat{\mathbb{P}}(\vec{k},\vartheta,t| \vartheta_0) = \frac{1}{2\pi} \sum_{\mu,\nu=-\infty}^\infty F_{\mu\nu}(\vec{k},t) \e^{-\i \mu \vartheta_0 + \i \nu \vartheta}
\end{align}
with the restricted ISF as
\begin{align}\label{eq:restricted_ISF}
F_\nu( \vec{k}, t| \vartheta_0 ) = \sum_{\mu=-\infty}^\infty F_{\mu\nu}(\vec{k},t) \e^{-\i \mu \vartheta_0} . 
\end{align}
The generalized ISF, Eq.~\eqref{eq:ISF2}, can be written as correlation function 
\begin{align}
F_{\mu\nu}(\vec{k},t) = \langle \rho_\mu(\vec{k},0)^* \rho_\nu(\vec{k},t) \rangle
\end{align}
of the fluctuating angle-resolved densities $\rho_\nu(\vec{k},t) = \exp(-\i \vec{k}\cdot \vec{R}(t) - \i \nu \Theta(t))$.  Therefore, $F_{\mu\nu}(\vec{k},t)$  inherits the usual properties of matrix-valued equilibrium correlation functions; for a complete characterization see e.g.~Ref.~\cite{Lang:JSM_2013:2013}. In particular, $F_{\mu\nu}(\vec{k},t) = F_{\mu\nu}(\vec{k},-t) = F_{\nu\mu}(\vec{k},t)^*$. 

The symmetries of the e.o.m. entail restrictions for  the propagator $\mathbb{P}(\vec{k}, \vartheta, t| \vartheta_0)$ and its Fourier decompositions $F_\nu(\vec{k}, t| \vartheta_0)$ and $F_{\mu\nu}(\vec{k},t)$, which can be used to reduce them to a particularly simple form.  We elaborate these for $F_{\mu\nu}(\vec{k},t)$. 
Applying a  rotation ${\sf R}_\chi$ by an angle $\chi$ of the laboratory frame results by statistical isotropy in  
\begin{align}\label{eq:rotinv}
F_{\mu\nu}(\vec{k},t) =\e^{\i (\nu-\mu) \chi} F_{\mu\nu} ({\sf R}_\chi^{-1} \cdot \vec{k}, t)
\end{align} 
implying that only the diagonal elements $\mu=\nu$ are rotationally invariant and depend solely on the magnitude of the scattering vector $k=|\vec{k}|$. In particular, $F_{00}(\vec{k},t) = F(k,t)$ corresponds to the conventional self-intermediate scattering function, Eq.~\eqref{eq:convISF}. Furthermore,  for vanishing wave number the $F_{\mu\nu}(0,t)$ are diagonal in the mode indices $\mu,\nu$.  

Since  rotating the initial orientation of the particle  by $\pi$ generates the same trajectory (with rotated orientation) for the same noise history,  the Fourier modes fulfill 
\begin{align}
F_{\mu\nu}(\vec{k},t) =  \langle \e^{-\i \vec{k} \cdot \Delta \vec{R}(t)} \e^{\i \mu (\Theta(0)+\pi) }\e^{-\i \nu (\Theta(t)+\pi) }   \rangle
\end{align} 
and correspondingly, the non-vanishing components of $F_{\mu\nu}(\vec{k},t)$ fulfill the selection rule that $\nu-\mu$ has 
to be an even number or, in other words, both mode indices are either even or odd. Then  Eq.~\eqref{eq:rotinv} yields, in particular, $F_{\mu\nu}(\vec{k},t) = F_{\mu\nu}(-\vec{k},t)$ for rotations of the laboratory coordinates by $\chi=\pi$. Since the dynamic 
evolution, Eqs.~\eqref{eq:LangevinAngle} and~\eqref{eq:LangevinPos}, is non-chiral, we also find  the symmetry $F_{\mu\nu}(\vec{k},t) = 
F_{-\mu, -\nu}((k_x,-k_y)^{\text{T}}, t) = F_{\mu \nu}((-k_x, k_y)^{\text{T}},t)^*$. 
Taking all these considerations together suggests choosing the scattering vector in the $x$-direction, since then  the quantity  $F_{\mu\nu}(k,t) :=F_{\mu\nu}(k \vec{e}_x,t)$ is  a real symmetric matrix in $\mu,\nu$, which is invariant under a simultaneous sign flip of the mode indices. If we impose that the initial orientation is along the $x$-direction, Eq.~\eqref{eq:restricted_ISF} reveals that also the restricted ISF $F_\nu(k, t | \vartheta_0=0) := F_\nu(k \vec{e}_x, t | \vartheta_0=0)$ is a real quantity.

\subsection{The Smoluchowski-Perrin equation}

By standard methods of stochastic calculus \cite{Gardiner:Stochastic_Methods:2009,Doi:Theory_of_Polymer_Dynamics:1986}  the Fokker-Planck equation for the time evolution of the conditional probability density $\mathbb{P}\equiv\mathbb{P}(\vec{r},\vartheta,t| \vartheta_{0})$ is derived from the corresponding Langevin equations.  For the case of anisotropic diffusion,  Eqs.~\eqref{eq:LangevinAngle} and \eqref{eq:LangevinPos}, we arrive at the Smoluchowski-Perrin equation~\cite{Perrin:JPRadium_7:1936,Doi:Theory_of_Polymer_Dynamics:1986}  
\begin{equation}\label{eq:Smoluchowski}
\partial_{t}\mathbb{P}=D_{\text{rot}}\partial_{\vartheta}^{2}\mathbb{P}+\nabla_{\mathbf{r}}\cdot\left( \textsf{D} \cdot\nabla_{\mathbf{r}}\right)\mathbb{P},
\end{equation}
where  $\nabla_{\mathbf{r}}$ denotes the spatial gradient and $\textsf{D}$ the translational diffusion tensor with  components $D_{ij}=D_{\parallel}u_{i}u_{j}+D_{\perp}\left(\delta_{ij}{-}u_{i}u_{j}\right)$.  
It is supplemented by  the initial condition $\mathbb{P}(\vec{r},\vartheta,t{=}0 | \vartheta_{0})=\delta(\vec{r})\delta_{2\pi}(\vartheta{-}\vartheta_{0})$, Eq.~\eqref{eq:Pini}. 
For anisotropic particles, $D_\parallel > D_\perp$, there is a non-trivial coupling between  the translational motion and the orientation $\vec{u} = (\cos\vartheta,\sin\vartheta)^{\text{T}}$ of the particle.

Translational invariance of the problem suggests performing a spatial Fourier transform, 
Eq.~\eqref{eq:Spatial_Fourier}, which readily leads to the e.o.m.\@ for the propagator $\hat{\mathbb{P}}\equiv \hat{\mathbb{P}}(\vec{k},\vartheta,t| \vartheta_0)$ in the Fourier domain 
\begin{equation}\label{eq:SmoluchowskiFourier}
\partial_{t}\hat{\mathbb{P}}=D_{\text{rot}}\partial_{\vartheta}^{2}\hat{\mathbb{P}}-D_{\perp}k^{2}\hat{\mathbb{P}}-\left[D_{\parallel}-D_{\perp}\right](\mathbf{u}\cdot\mathbf{k})^{2}\hat{\mathbb{P}},
\end{equation}
subject to the initial condition $\hat{\mathbb{P}}(\vec{k},\vartheta,t{=}0| \vartheta_0) = \delta_{2\pi}(\vartheta{-}\vartheta_0)$, Eq.~\eqref{eq:Pkini}. 
This equation and its generalizations have already been solved formally and numerically for passive and active anisotropic Brownian particles 
in two~\cite{Munk:EPL_85:2009,Munk:PhD_thesis:2008,Kurzthaler:SM_13:2017,Kurzthaler:PRL_121:2018} and 
three dimensions \cite{Aragon:JCP_82:1985,Berne:Dynamic_Light_Scattering:2000, Otto:JCP_124:06, Leitmann:PRL_117:2016,Kurzthaler:SciRep_6:2016, Mandal:PRL_125:2020}. 
Yet, neither solutions for  $F_{\mu}(\mathbf{k},t\vert\vartheta_{0})$ nor for $F_{\mu\nu}(k,t)$ have been investigated so far (except for 
\cite{Munk:PhD_thesis:2008}).

Following the strategy outlined in Refs.~\cite{Munk:EPL_85:2009,Munk:PhD_thesis:2008} for a formally exact solution,  
the direction of the wavevector $\mathbf{k}=k \mathbf{e}_{x}$ is chosen such that the Fourier modes $F_{\mu\nu}(k,t)$ become real. Then 
the previous equation for $\hat{\mathbb{P}}(k,\vartheta, t| \vartheta) := \hat{\mathbb{P}}(k\vec{e}_x, \vartheta, t| \vartheta_0)$ simplifies to
\begin{equation}\label{eq:SmoluchowskiFourier2}
\partial_{t}\hat{\mathbb{P}}=\left[D_{\text{rot}}\partial_{\vartheta}^{2}-k^2\left(\bar{D}+ \frac{\Delta D}{2} \cos(2\vartheta)\right)\right]\hat{\mathbb{P}} .
\end{equation}
We seek solutions of  Eq.~\eqref{eq:SmoluchowskiFourier2} using the separation ansatz $\hat{\mathbb{P}} = z(\vartheta) \exp(-\lambda t)$.
Then  angular eigenfunctions $z(\vartheta)$ solve  the eigenvalue problem 
\begin{equation}\label{eq:Mathieuequ}
\left[\frac{\diff^2}{\diff \vartheta^2}+a-2q\cos(2\vartheta)\right]z(\vartheta)=0
\end{equation}
resulting in the standard form of the Mathieu equation~\cite{Nist:website_Mathieu:2010}
with the eigenvalues $a=(\lambda-k^{2}\bar{D})/D_{\text{rot}}$ and 
the deformation parameter $q=k^{2}\Delta D/4D_{\text{rot}}\geq 0$  quantifying the ratio of the
rotational relaxation time to the anisotropic diffusional relaxation time  at length scale $2\pi/k$.  

The general solution of Eq.~\eqref{eq:Mathieuequ} is expressed as a linear superposition of even $\text{ce}_{n}(\vartheta, q), n\in \mathbb{N}_0$, and odd $\text{se}_{n}(\vartheta, q), n\in \mathbb{N}$, Mathieu functions~\cite{Nist:Handbook_of_Mathematical_Functions:2010, Nist:website_Mathieu:2010}  with corresponding eigenvalues $a_{n}(q)$ and $b_{n}(q)$ \cite{Nist:Handbook_of_Mathematical_Functions:2010, Nist:website_Mathieu:2010}. These Mathieu functions reduce to  $\text{ce}_0(\vartheta, 0) = 1/\sqrt{2}$ and  $\text{ce}_n(\vartheta, 0) = \cos(n \vartheta),  \text{se}_n(\vartheta, 0) = \sin(n \vartheta), n\in \mathbb{N}$, for vanishing deformation parameter $q=0$ and the corresponding eigenvalues reduce to $a_n(0) =  b_n(0)= n^2$. 
The Mathieu functions constitute a complete set of orthogonal eigenfunctions normalized as
\begin{subequations}
\begin{align}
\int_{0}^{2\pi}\diff \vartheta\, \text{ce}_{n}(\vartheta, q)\text{ce}_{m}(\vartheta, q)&=\pi\delta_{nm}, \\
\int_{0}^{2\pi}\diff \vartheta\, \text{ce}_{n}(\vartheta, q)\text{se}_{m}(\vartheta, q)&=0
\end{align}
\end{subequations}
and similar for $\text{se}_{n}(q,\vartheta_{0})$. Hence, the full solution of Eq.~\eqref{eq:SmoluchowskiFourier2} is
obtained~\cite{Munk:PhD_thesis:2008,Munk:EPL_85:2009} as
\begin{align}\label{eq:expansion}
&\hat{\mathbb{P}}(k,\vartheta,t\vert\vartheta_{0})= \nonumber \\ 
&\qquad\frac{\text{e}^{-k^{2}\bar{D}t}}{\pi}\sum_{n=0}^{\infty}\left[\text{ce}_{n}(\vartheta, q)\text{ce}_{n}(\vartheta_0, q)\text{e}^{-a_{n}(q)D_{\text{rot}}t}\right. \nonumber \\
&\left.\qquad\qquad\quad+\text{se}_{n}(\vartheta, q)\text{se}_{n}(\vartheta_0, q)\text{e}^{-b_{n}(q)D_{\text{rot}}t}\right],
\end{align}
where we define $se_{0}(\vartheta, q):=0$ to let the sum start at $n=0$.
By completeness of the eigenfunctions, this representation indeed fulfills the initial condition $\hat{\mathbb{P}}(k,\vartheta, t=0|\vartheta_{0})=\delta_{2\pi}(\vartheta{-}\vartheta_{0})$ for $t=0$. Furthermore the eigenvalues for $q>  0$ are ordered with increasing $n$~\cite{Nist:Handbook_of_Mathematical_Functions:2010, Nist:website_Mathieu:2010}
\begin{align}
 a_{0}(q)<b_{1}(q)<a_{1}(q)<b_{2}(q)<a_{2}(q)<\ldots 
\end{align}
indicating the convergence of the series expansion for any $t>0$. Thus, Eq.~\eqref{eq:expansion} can be efficiently evaluated numerically, which is discussed in more  detail in Appendix \ref{app:numMathieu}.

\subsection{Restricted and generalized self-intermediate scattering functions}

The restricted ISF $F_\nu(k,t| \vartheta)$ is obtained by a decomposition of the final angle $\vartheta$ into Fourier modes, Eq.~\eqref{eq:restricted_ISF}.  We distinguish the cases for zero mode index
\begin{align}\label{eq:ISFana0}
F_{0}(k,t\vert\vartheta_{0})&=2\,\text{e}^{-k^2\bar{D}t} \nonumber \\
&\times \sum_{n=0}^{\infty}\text{ce}_{2n}(\vartheta_{0}, q)\text{e}^{-a_{2n}(q)D_{\text{rot}}t}A_{0}^{2n}(q)
\end{align}
and non-vanishing mode index $\nu\neq 0$
\begin{align}\label{eq:ISFana1}
F_{\nu}(k,t\vert\vartheta_{0})&=\text{e}^{-k^2\bar{D}t}\sum_{n=0}^{\infty}\left[\text{ce}_{n}(\vartheta_{0}, q)
\text{e}^{-a_{n}(q)D_{\text{rot}}t}A_{\nu}^{n}(q)\right. \nonumber \\
&\left.-\text{i}\,\text{se}_{n}(\vartheta_{0}, q)\text{e}^{-b_{n}(q)D_{\text{rot}}t}B_{\nu}^{n}(q)\right].
\end{align}
The real expansion coefficients $A^{n}_\nu(q), B^{n}_\nu(q)$ arise by the  Fourier cosine and sine representation of the Mathieu functions  as summarized in Eqs.~\eqref{eq:Fourier1}-\eqref{eq:Fourier4} in the Appendix~\ref{app:numMathieu}. Since the Mathieu functions are either $\pi$-periodic or $\pi$-antiperiodic, the Fourier coefficients $A^{n}_\nu(q), B^{n}_\nu(q)$ are non-vanishing if both indices have the same parity, which implies that 
the sum in Eq.~\eqref{eq:ISFana1} can be restricted to even (odd) numbers  if $\nu$ is even (odd).  For negative mode indices $\nu$, they obey the transformation rule $\left(A_{-\nu}^{n}(q),B_{-\nu}^{n}(q)\right)^{\text{T}}\mapsto \left(A_{\nu}^{n}(q),-B_{\nu}^{n}(q)\right)^{\text{T}}$. 
In general, the restricted generalized ISF are complex quantities except for $F_0(k,t|\vartheta_0)$, but become real for the special case of $\vartheta_0=0$ with $F_{-\nu}(k,t\vert\vartheta_{0}{=}0)=F_{\nu}(k,t\vert\vartheta_{0}{=}0)$.

Upon further integration over the initial angle the generalized  ISF, Eq.~\eqref{eq:ISF2},
 for zero mode indices reduces  to the conventional ISF~\cite{Munk:EPL_85:2009} 
\begin{equation}\label{eq:F00}
 F_{00}(k,t)=2\,\text{e}^{-k^2\bar{D}t}\sum_{n=0}^{\infty}\text{e}^{-a_{2n}(q)D_{\text{rot}}t}\left[A_{0}^{2n}(q)\right]^2 .
\end{equation}
 For the case $\mu=0$ and even $\nu\neq 0 $  or vice versa we find
\begin{equation}\label{eq:F0nu}
F_{0\nu}(k,t)=\text{e}^{-k^2\bar{D}t}\sum_{n=0}^{\infty}\text{e}^{-a_{2n}(q)D_{\text{rot}}t}A_{0}^{2n}(q)A_{\nu}^{2n}(q).
\end{equation}
Last, for both mode indices nonzero and  $\mu-\nu$ an even number it assumes  the following form
\begin{align}\label{eq:Fmunu}
F_{\mu\nu}(k,t)&=\frac{\text{e}^{-k^2\bar{D}t}}{2}\sum_{n=0}^{\infty}
\left[\text{e}^{-a_{n}(q)D_{\text{rot}}t}A_{\mu}^{n}(q)A_{\nu}^{n}(q)\right. \nonumber \\
&\left.+\,\text{e}^{-b_{n}(q)D_{\text{rot}}t}B_{\mu}^{n}(q)B_{\nu}^{n}(q)\right].
\end{align}
For any set of complex numbers $y_\mu$ the quantity $f(t) :=\sum_{\mu\nu} y_\mu^* F_{\mu\nu}(k,t) y_\nu \geq 0$ is a completely monotone function, i.e. $(-1)^l \diff^l f(t)/\diff t^l \geq 0$. This property can be inferred as general consequence for matrix-valued autocorrelation functions in pure relaxational dynamics, for further properties see e.g. Ref.~\cite{Lang:JSM_2013:2013}. In particular, the diagonal elements $F_{\mu\mu}(k,t)$ are  completely monotone. The explicit solution shows that the spectrum of relaxation rates consists of isolated points, which implies that the long-time decay is dominated by the lowest eigenvalue. The admixture of the higher eigenvalues decays exponentially fast and its relative rate is determined by the spectral gap from the lowest to the next-lowest eigenvalue contributing to the expansion. 
However, numerical results reveal that the differences to a pure exponential decay remain small, but the admixture of several exponentials becomes clearly visible on  double-logarithmic scales~\cite{Munk:EPL_85:2009}.

For reference, we also provide the trivial case of vanishing TR coupling ($D_\parallel = D_\perp$). Then the  Smoluchowski-Perrin equation after a spatial Fourier transform, Eq.~\eqref{eq:SmoluchowskiFourier2}, becomes readily solvable, in particular, it is independent of the direction of the scattering vector $\vec{k}$.  
The propagator for isotropic diffusion can then be evaluated as
\begin{align}\label{eq:Ptrivial_Fourier}
\hat{\mathbb{P}}^{\text{Iso}}(k,\vartheta,t|\vartheta_{0}) &=\frac{\text{e}^{-k^{2}\bar{D}t}}{2\pi}\sum_{\nu=-\infty}^{\infty}\text{e}^{-\nu^{2}D_{\text{rot}}t}\text{e}^{\text{i}\nu\left(\vartheta-\vartheta_{0}\right)} \nonumber \\
&= \frac{\e^{- k^2 \bar{D} t}}{ 2 \pi } \theta_3\left(\frac{\vartheta-\vartheta_0}{2}, \e^{-D_{\text{rot}} t}\right) ,
\end{align}
where $\theta_3(z,q)$ is a (Jacobi) elliptic theta function~\cite{Nist:Handbook_of_Mathematical_Functions:2010, Nist:website_Theta:2010}. 
In particular, the completeness of the Fourier modes shows that  the initial condition $\hat{\mathbb{P}}^{\text{Iso}}(k,\vartheta,0\vert\vartheta_{0})=\delta_{2\pi}(\vartheta{-}\vartheta_{0})$ is satisfied. 
Alternatively, one observes that the e.o.m., Eqs.~\eqref{eq:LangevinAngle}-\eqref{eq:LangevinPos}, 
decouple and $(\vec{R}(t), \Theta(t))_{t\in\mathbb{R}}^{\text{T}}$ are 
independent Gaussian processes. Therefore the propagator is  obtained  by folding the final angle back to the interval $[0,2\pi)$
\begin{align}\label{Ptrivial_sum}
\hat{\mathbb{P}}^{\text{Iso}}(k,\vartheta,t | \vartheta_0) &=  \frac{\e^{- k^2 \bar{D} t}}{\sqrt{4\pi D_{\text{rot}} t}} \nonumber \\
&\times\sum_{n=-\infty}^\infty \exp\left( - \frac{(\vartheta -\vartheta_0 + 2\pi n)^2}{4D_{\text{rot}} t }\right),
\end{align}
also referred to as wrapped normal distribution.  The representation in Eq.~\eqref{eq:Ptrivial_Fourier} is  the Fourier decomposition of the infinite sum in Eq.~\eqref{Ptrivial_sum}. 
For the restricted ISF this implies
\begin{equation}\label{eq:ISFiso1}
F_{\nu}^{\text{Iso}}(k,t|\vartheta_{0})= \e^{-k^2 \bar{D} t} \e^{-\nu^2 D_{\text{rot} } t} \e^{-\i \nu \vartheta_0 }
\end{equation}
and in particular it is a real function for $\vartheta_{0}=0$. Then by Eq. \eqref{eq:ISF2} the generalized ISF evaluates to
\begin{equation}\label{eq:ISFiso2}
F_{\mu\nu}^{\text{Iso}}(k,t)= \e^{-k^2 \bar{D} t} \e^{-\nu^2 D_{\text{rot} } t} \delta_{\mu\nu}
\end{equation}
and becomes diagonal in the mode indices.
Moreover, the conventional self-intermediate scattering function for an isotropic particle becomes $F^{\text{Iso}}(k,t) := F^{\text{Iso}}_{00}(k,t) = \exp{(-k^2 \bar{D} t)}$ as it should.

\subsection{Exact expressions for the low-order moments via perturbation theory}

In this subsection we derive  the first non-trivial moments of the displacements $\Delta \vec{R}(t) = (\Delta X(t), \Delta Y(t))^{\text{T}}$ correlated 
with the  angular modes $\e^{-\i \nu \Theta(t)}$  for both the equilibrium dynamics and  for a  fixed initial angle $\Theta(0) = \vartheta_0$. 
Moments up to fourth order have been derived earlier  by Han \textit{et al.}~\cite{Han:Science_314:2006} 
by
directly investigating the Langevin equations. Here we rederive their expressions using the Smoluchowski-Perrin equation, Eq. \eqref{eq:SmoluchowskiFourier}.

Since the restricted ISF $F_\nu(\vec{k},t|\vartheta_0)$ can be viewed as a characteristic (or moment-generating) function,  these low-order moments are obtained by taking suitable derivatives with respect to the scattering vector $\vec{k}= (k_x,k_y)^{\text{T}}$,
\begin{align}\label{eq:moments}
&\langle\Delta X(t)^{n}\Delta Y(t)^{m} \e^{-\i \nu \Theta(t)} \rangle_{\vartheta_{0}}= \nonumber \\
&\qquad\qquad(-\i)^{n+m}\frac{\partial^{n}}{\partial k_x^{n}}\frac{\partial^{m}}{\partial k_y^{m}}F_{\nu}(\vec{k},t\vert\vartheta_{0})\bigg\vert_{\vec{k}=0}.
\end{align}
The equilibrium dynamics is recovered by averaging over the initial angle. In principle, one can use the asymptotic expansions of the Mathieu function~\cite{Nist:Handbook_of_Mathematical_Functions:2010,Nist:website_Mathieu:2010}  to evaluate the derivatives directly  
as has been elaborated  in Ref.~\cite{Kurzthaler:PRE_E95:2017} for the case of a semiflexible polymer. Here we provide  a systematic method to generate moments of arbitrary order.

The Smoluchowski-Perrin equation after spatial Fourier transform, Eq.~\eqref{eq:SmoluchowskiFourier}, contains the scattering vector in the operator acting on the propagator. The strategy is  to solve  the equation perturbatively for small scattering vectors. In a first step we reformulate the problem in terms of  the restricted ISF by multiplying the Smoluchowski-Perrin equation by $\e^{-\i \nu \vartheta}$ and integrating over the final angle $\vartheta$. Then one finds the  set of coupled ordinary differential equations
\begin{equation}\label{eq:coupled_ode}
\!\!\left[\partial_t + \nu^2 D_{\text{rot}}\right] F_\nu(\vec{k},t| \vartheta_0)=-\!\sum_{\sigma=-\infty}^{\infty}\! V_{\nu\sigma} F_\sigma(\vec{k},t |\vartheta_0 ) ,
\end{equation} 
where the propagator on the r.h.s.\@ is expressed in terms of the restricted ISF. The integral on the r.h.s.\@ can be performed and defines the matrix elements of the perturbation with respect to the Fourier basis
\begin{align}
V_{\nu\sigma}&=\int_{0}^{2\pi}\textrm{d}\vartheta\,\e^{-\i \nu\vartheta}\left[D_{\perp}+\Delta D\left(\vec{u}\cdot\vec{k}\right)^{2}\right]\e^{\i \sigma\vartheta}\nonumber \\
&=\bar{D} k^2 \delta_{\nu,\sigma} + \frac{\Delta D}{2}\left[k_{-}^2 \delta_{\nu,\sigma+2} + k_{+}^2 \delta_{\nu,\sigma-2}\right]
\end{align}
with the wavenumbers in the spherical basis $k_\pm = (k_x \pm \i k_y)/\sqrt{2}$. In particular, the matrix elements of the perturbation $V_{\nu\sigma}$ are quadratic in the scattering vector.  
The solutions to vanishing scattering vector fulfilling the proper initial condition 
are found to be $F_\nu(\vec{k}{=}0,t| \vartheta_0) = \exp(- \nu^2 D_{\text{rot}} t - \i \nu \vartheta_0)$. 
Then we can generate the low-scattering-vector expansion of the restricted ISF by  solving Eq.~\eqref{eq:coupled_ode} iteratively: Substituting the $n^\text{th}$-order solution on the r.h.s.\@ and solving for the l.h.s.\@ one finds the $(n+1)^{\text{th}}$-order solution. 
Up to fourth order in the wave vector  we obtain
\begin{align}\label{eq:F_nu_second_order}
&F_{\nu}(\vec{k},t\vert\vartheta_{0})=\e^{-\nu^{2}D_{\text{rot}}t}\e^{-\i \nu\vartheta_{0}} \left[\vphantom{\frac{\Delta D}{4}} 1-\bar{D}k^{2}t\right. \nonumber \\
&\qquad\quad\left.-\frac{1}{2}\Delta D\left(\tau_{4-4\nu}(t)k_{-}^{2}\e^{2\i \vartheta_{0}}+\tau_{4+4\nu}(t)k_{+}^{2}\e^{-2\i \vartheta_{0}}\right)\right. \nonumber \\
&\qquad\quad\left.+\frac{1}{2}\bar{D}^{2}k^{4}t^{2}+\frac{(\Delta D)^{2}}{16}k^{4}\left(T_{4+4\nu}^{4-4\nu}(t)+T_{-4-4\nu}^{4+4\nu}(t)\right)\right. \nonumber \\
&\qquad\quad\left.+\frac{1}{2}\bar{D}\Delta D k^{2}\left(T_{4-4\nu}^{0}(t)+T_{0}^{4-4\nu}(t)\right)k_{-}^{2}\e^{2\i \vartheta_{0}}\right. \nonumber \\
&\qquad\quad\left.+\frac{1}{2}\bar{D}\Delta D k^{2}\left(T_{4+4\nu}^{0}(t)+T_{0}^{4+4\nu}(t)\right)k_{+}^{2}\e^{-2\i\vartheta_{0}}\right. \nonumber \\
&\qquad\quad\left.+\frac{1}{4}(\Delta D)^{2}\left(T_{12-4\nu}^{4-4\nu}(t)k_{-}^{4}\e^{4\i \vartheta_{0}}+T_{12+4\nu}^{4+4\nu}(t)k_{+}^{4}\e^{-4\i \vartheta_{0}}\right)\right. \nonumber \\
&\qquad\quad\left.+\mathcal{O}(k^{6})\vphantom{\frac{\Delta D}{4}}\right] ,
\end{align}
where we follow Ref.~\cite{Han:Science_314:2006} and abbreviate the integrals
\begin{align}\label{eq:tau_n}
\tau_n(t) :=& \int_0^t \, \e^{- n D_{\text{rot}} s} \diff s \nonumber \\
=& \begin{cases}
t, & n=0 \\
\left[1- \exp(- n D_{\text{rot}} t)\right] /(n D_{\text{rot}}), & n\neq 0
\end{cases},
\end{align}
together with
\begin{align}
T_n^m(t) := \int_0^t \e^{-m D_{\text{rot}} s}\tau_n(s) \diff s .
\end{align}
Extending the perturbation scheme to higher orders in the scattering vector is straightforward but tedious. 
From Eq.~\eqref{eq:moments}
we recover the restricted mean-square displacements~\cite{Han:Science_314:2006}
\begin{subequations}
\begin{align}
\langle\Delta X(t)^{2}\rangle_{\vartheta_{0}}&=2\bar{D}t+\Delta D\,\tau_{4}(t)\cos(2\vartheta_{0}), \label{eq:MSDComp1}\\
\langle\Delta Y(t)^{2}\rangle_{\vartheta_{0}}&=2\bar{D}t-\Delta D\,\tau_{4}(t)\cos(2\vartheta_{0}). \label{eq:MSDComp2}
\end{align}
\end{subequations}
More complicated correlations, such as the fourth moments $\langle \Delta X(t)^4 \rangle_{\vartheta_0} $ and $\langle \Delta Y(t)^4 \rangle_{\vartheta_0} $ as discussed in Ref.~\cite{Han:Science_314:2006}, are also encoded in Eq.~\eqref{eq:F_nu_second_order}.

\section{Materials and methods}\label{sec:MM}

\subsection{Sample preparation}

An aqueous suspension of polystyrene spheres with negatively charged sulfonated chain ends and a diameter of $\sigma = 2\,\text{\textmu}\text{m}$ (Duke Scientific) is  dialyzed against ultrapure water to eliminate any surfactant that might be present.
Then dimers are formed following a previously developed procedure~\cite{Castillo:PhysRevE_78:2008}.
The suspension is filled into a dialysis bag and placed in a $500\,\text{mM}$ NaCl solution to induce aggregation of the particles.
 After six minutes the aggregation is stopped by dialysis against ultra pure water (resistivity $18\,\text{M}\Omega\text{cm}$).
To create a permanent contact between touching spherical particles, the suspension 
is kept for $15\,\text{min}$ at a temperature close to the melting temperature of polystyrene ($104^\circ\,\text{C}$). 
This sinters the touching particles and results in stable dimers but also other multimers.
The suspension is cooled down to room temperature ($24^\circ\,\text{C}$).
The dimers are enriched by centrifugation in the density gradient of a sugar solution.
Subsequently the sugar is removed by dialysis.
Finally, surfactant (SDS with a concentration corresponding to the cmc) is added to avoid further aggregation.
The investigated sample consists of dimers and individual spheres  with a number fraction of dimers of $0.734$ and a total surface fraction of 0.1425 (Sample d in~\cite{Sarmiento:PhysRevE_94:2016}).
It is assumed that the presence of  individual spheres does not significantly change the dynamics of the dimers. We have tested this assumption by analyzing trajectories for dimers in a sample with a
lower total surface fraction of $0.09$. The results are consistent, which suggests that particle interactions
can be neglected. Subsequently, we only discuss the data for $0.1425$, which result in smaller uncertainties.

For the measurements, the sample is confined between a microscope slide and a coverslip with the gap sealed with epoxy resin.
The sample contains trace amounts of larger polystyrene spheres with negatively charged sulfonated chain ends and a diameter of $2.98\,\text{\textmu m}$ (Duke Scientific) that serve as spacers and hence define the gap height of the sample cell.
Due to the very narrow gap, any excursion perpendicular to the wall is greatly limited. 
This results in a quasi-2D geometry.
The particles are kept in the sample cell for at least $2\,\text{h}$ at room temperature before measurements are performed.

\subsection{Video microscopy and particle tracking}

Measurements are performed at room temperature $(24.0 \pm 0.1)^\circ\,\text{C}$.
Due to the quasi-2D geometry the particles are imaged by conventional optical microscopy.
An inverted bright field microscope (Olympus, IX$71$) was used with a $40\times$ objective (Olympus, LC$\,$Plan$\,$FI, numerical aperture $0.6$) and a polychromatic light source.
Images are recorded using a CCD camera with $640 \times 480$ pixels (WAT-902H3 Supreme) and directly stored on a DVD and then converted to frames using frame grabber decoding software.
The field of view is $85.1 \times 63.8\,\text{\textmu}\text{m}^2$ and the pixel pitch $0.133\,\text{\textmu}\text{m/px}$.
During a measurement with a frame rate of $30$~fps and lasting $20\,\text{min}$, a series of $36\,000$ images is recorded.
We performed three such measurements.
Only dimers that are tracked during the whole measurement time are considered.
This results in $N_\text{t} = 147$ trajectories of duration $T = 20\,\text{min}$ each.

The positions of individual spheres relative to the laboratory frame $\mathbf{R}_{\text{s}}(t_n)=(X_{\text{s}}(t_n),Y_{\text{s}}(t_n))^{\text{T}}$ are extracted from the recorded images taken at times $t_n = n \Delta t$ with
$n\in\{0,\ldots,N-1\}$  and the time between two images $\Delta t = (1/30)\,\text{s}$ given by the inverse frame rate.
(Artefacts due to dynamic errors are described in Appendix~\ref{app:Exposuretime}.)
The positions are connected to yield trajectories using a custom-written IDL program based on a standard algorithm \cite{Crocker:JCIS_179:1996}.
Dimers are identified by determining the distance between two neighboring spheres as a function of time.
If two spheres stay closer than $1.25\,\sigma$ during the whole measurement, they are considered to form a dimer.
The centers of two particles forming a dimer are found to be on average $(0.90 \pm 0.04)\,\sigma$ apart.
This is closer than a particle diameter and a consequence of the sintering process.

Based on the positions of the particles forming the dimer, the position of the center of mass in the laboratory frame $\vec{R}^{(j)}(t_{n})=(X^{(j)}(t_{n}), Y^{(j)}(t_{n}))^{\text{T}}$ and the instantaneous orientation $\Theta^{(j)}(t_{n})$ of all dimers $j$ are determined.
Between two images, the dimers typically move by approximately $0.07\,\text{\textmu{}m} = 0.035 \,\sigma$ and rotate by approximately $0.04\,\text{rad}\approx 2.3^{\circ}$.
These distances and rotations are small enough that particles can be followed between consecutive images and trajectories can be assigned unambiguously.
The time evolutions of particle positions  yield the trajectories of the dimers $\vec{R}^{(j)}(t_n)$ and the absolute angles $\Theta^{(j)}(t_{n})$ with 
the total number of turns calculated by  concatenating the rotations. 
Following \cite{Crocker:JCIS_179:1996, Michalet:PRE_82:2010}, the uncertainty of the location of an individual particle is estimated to be 
$\epsilon(X_\text{s}) = \epsilon(Y_\text{s})  = 0.029\,\text{\textmu}\text{m}$. 
Hence the uncertainty of the location of the center of mass of a dimer is
$\epsilon(X) = \epsilon(Y) = \epsilon(X_\text{s})/\sqrt{2}$ and
$\epsilon(\vec{R}) = (\epsilon^2(X))^{1/2} + \epsilon^2(Y))^{1/2} = 0.03\,\text{\textmu}\text{m}$. 
 Moreover, this implies an uncertainty of the dimer orientation 
$\epsilon(\Theta) = 2 \epsilon(X)/(0.9\,\sigma) = 0.02\,\text{rad} = 1.15^\circ$ where $0.9\,\sigma$ is the observed average distance between two spheres forming a dimer.
The transformation from the laboratory frame to the particle frame is described in Appendix~\ref{app:LabPartframe} and the calculation of the correlation functions from the dimer trajectories and orientations as well as their uncertainties are given in Appendix~\ref{app:Correlation_functions}.

\section{Results and discussion}\label{Sec:Results}

\begin{figure}[htbp]
\centering
\includegraphics[width=0.4\textwidth]{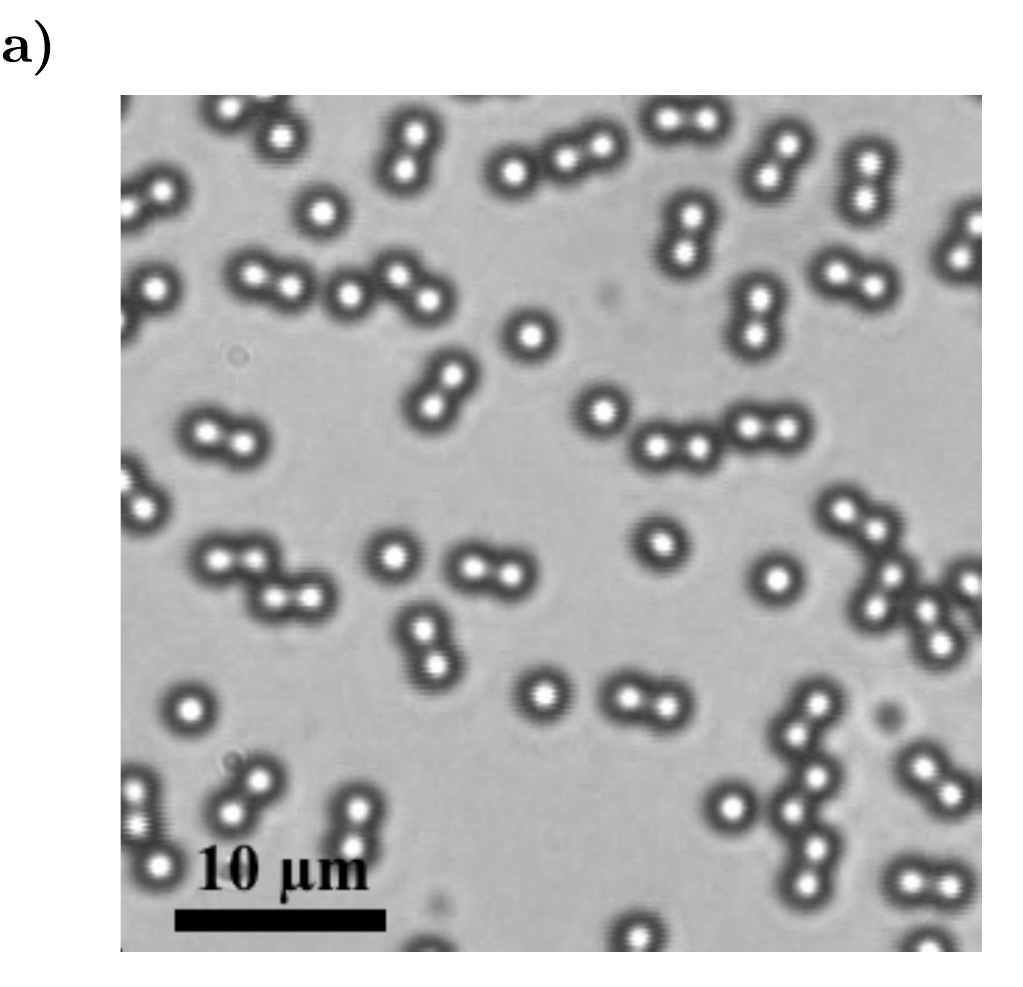}\\ \hspace*{-0.63cm}
\includegraphics[width=0.55\textwidth]{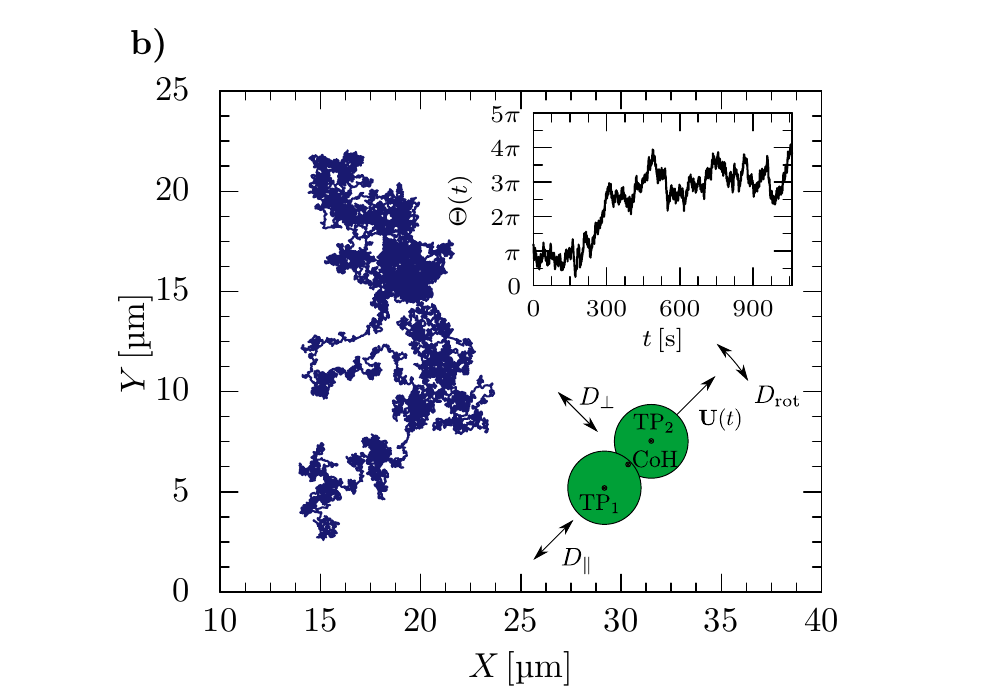}
\caption{\textbf{a}) Optical microscopy image of the sample representing only part of the field of view. The scale bar corresponds to $10\,\text{\textmu}\text{m}$.
\textbf{b)} Center of mass trajectory  of a single dimer in the laboratory frame lasting  $20\,\text{min}$ and obtained with a frame rate of $30$~fps (blue line). The inset shows the corresponding orientation $\Theta(t)$ for a subset of the  data points.  A schematic of the dimer is depicted, where the center of hydrodynamic stress (CoH)  coincides with the center of mass (CoM).  The latter is calculated from the two  centers of the particles forming the dimer, indicated by $\text{TP}_{1}$ and $\text{TP}_{2}$. The three modes of transport are shown together with the corresponding transport coefficients, the rotational self-diffusion coefficient $D_\text{rot}$ and the self-diffusion coefficients parallel $D_{\parallel}$ and perpendicular $D_{\perp}$ to the dimer orientation $\vec{U}(t)$.}
\label{fig:plottraj1}
\end{figure}

The dimers are followed by bright-field microscopy (Fig.~\ref{fig:plottraj1}a).
A quantitative analysis of the videos yields their positions $\vec{R}^{(j)}(t_{n})=(X^{(j)}(t_{n}), Y^{(j)}(t_{n}))^{\text{T}}$ and orientations $\Theta^{(j)}(t_{n})$, where at each time $t_n$ an image is taken,
They are combined to yield trajectories $(\vec{R}(t_n), \Theta(t_n))^{\text{T}}$ (Fig.~\ref{fig:plottraj1}b).

In the following, we compare the experimental results obtained from analyzing the trajectories (Sec.~\ref{sec:MM} and Appendix~\ref{app:Correlation_functions}) to the theoretical predictions  of Sec.~\ref{Sec:Theory} with particular emphasis on the effects of the TR coupling. We shall need from Sec.~\ref{Sec:Theory} only the results for the generalized intermediate scattering functions, Eqs.~\eqref{eq:F00}, \eqref{eq:F0nu} and \eqref{eq:Fmunu}, as well as their isotropic counterparts, Eqs.~\eqref{eq:ISFiso1} and \eqref{eq:ISFiso2}, and the expressions for the restricted mean-square 
displacements, Eqs.~\eqref{eq:MSDComp1} and  \eqref{eq:MSDComp2}.
However, first we check whether the experimentally determined center of mass (CoM) matches with the center of hydrodynamic stress (CoH) on which the theoretical predictions are based.

\subsection{Center of mass and center of hydrodynamic stress}
\label{CoM-CoH}

From the positions of the particles forming the dimer, the center of mass (CoM)  and the orientation of the dimer are calculated.
Although dimers are symmetric in principle, the actual constituent particles may differ in size due to polydispersity. Therefore, 
the determined CoM might not match the CoH for which we have developed the formalism. Therefore we check for possible differences.

\begin{figure}[htbp]
\centering
\includegraphics[width=0.5\textwidth]{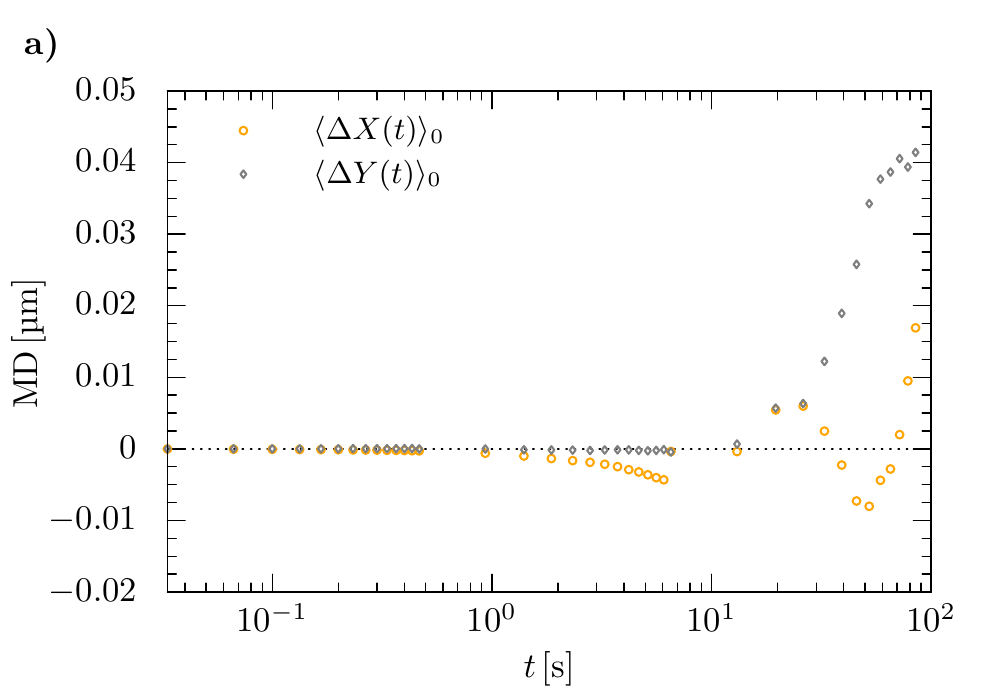}
\includegraphics[width=0.5\textwidth]{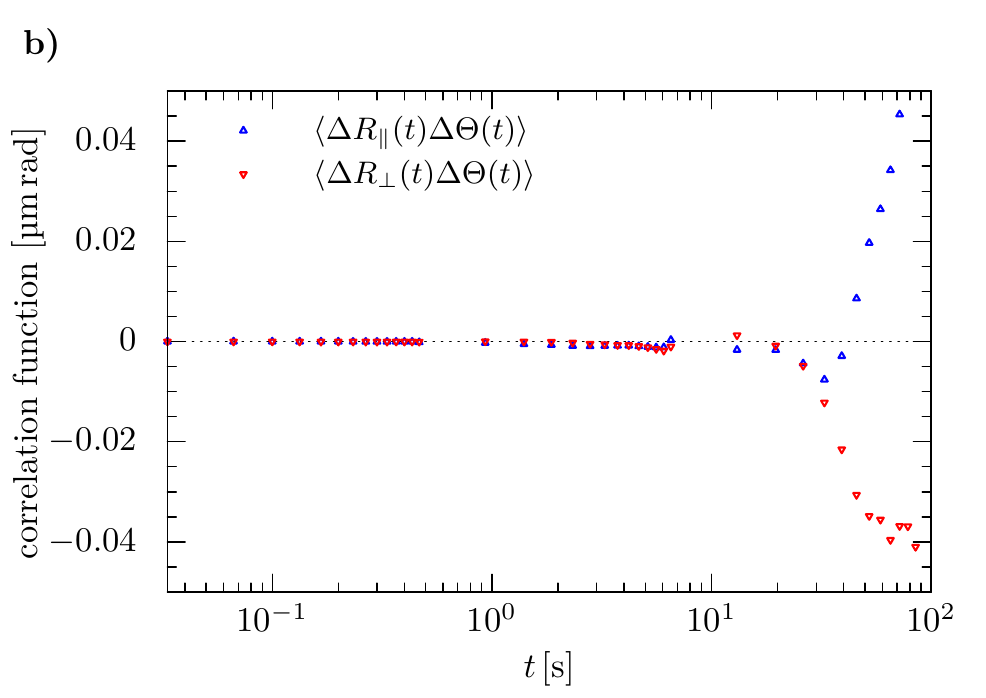}
\caption{ \textbf{a)} Restricted mean displacements (MD) $\langle\Delta X(t)\rangle_{0}$ and $\langle\Delta Y(t)\rangle_{0}$ of dimers initially oriented along the $x$-direction and \textbf{b)}  correlations between the displacements parallel or perpendicular to 
the principal axis and the angular increment, $\langle \Delta R_\parallel(t) \Delta \Theta(t) \rangle$ and $\langle \Delta R_\perp(t) \Delta \Theta(t)\rangle$, respectively.}
\label{fig:plotcorCoH}
\end{figure}

First, based on the experimental data the mean displacements $\langle\Delta X(t)\rangle_{0}$ and $\langle\Delta Y(t)\rangle_{0}$
are calculated with $\Delta X(t)=X(t{+}t_{0})-X(t_{0})$ and $\Delta Y(t)=Y(t{+}t_{0})-Y(t_{0})$,
 given that the initial orientation is in the $x$-direction ($\Theta(0)=\vartheta_{0}=0$). Here the brackets $\langle\ldots\rangle$ indicate a time and ensemble average over all trajectories. By rotating the trajectories accordingly, each data point can be chosen as an initial point with orientation $\vartheta_{0}=0$ thereby yielding good statistics also for these restricted averages. (We refer to Appendix~\ref{app:Correlation_functions} for details.)
 If the CoM differs from the CoH the restricted mean displacements exhibit  non-trivial short time dynamics~\cite{Chakrabarty:Langmuir_30:2014, Chakrabarty:SM_12:2016}. However, our data shows virtually no drift in the $y$-direction and only a very small one of approximately $5\,\text{nm}$ in the $x$-direction up to lag times of $10\,\text{s}$ beyond which the data is dominated by noise (Fig.~\ref{fig:plotcorCoH}\,a). For comparison,  typical displacements during this lag time are approximately $1\,\text{\textmu{}m}$. 

Second, differences between the CoM and the CoH also yield nontrivial correlations between displacements parallel or perpendicular to the principle axis and angular increments in the particle frame of reference, $\langle\Delta R_{\parallel}(t)\Delta\Theta(t)\rangle$ and $\langle\Delta R_{\perp}(t)\Delta\Theta(t)\rangle$ with $\Delta\Theta(t)=\Theta(t{+}t_{0})-\Theta(t_{0})$. The data (Fig.~\ref{fig:plotcorCoH}\,b) show no indication of such effects at lag times where the data is  not significantly  affected by noise.

 These results indicate that in our experiments the CoM and CoH match within the experimental uncertainties, as  expected for anisotropic particles of high symmetry such as dimers or ellipsoids~\cite{Han:Science_314:2006}.

\begin{figure}[htbp]
\centering
\includegraphics[width=0.5\textwidth]{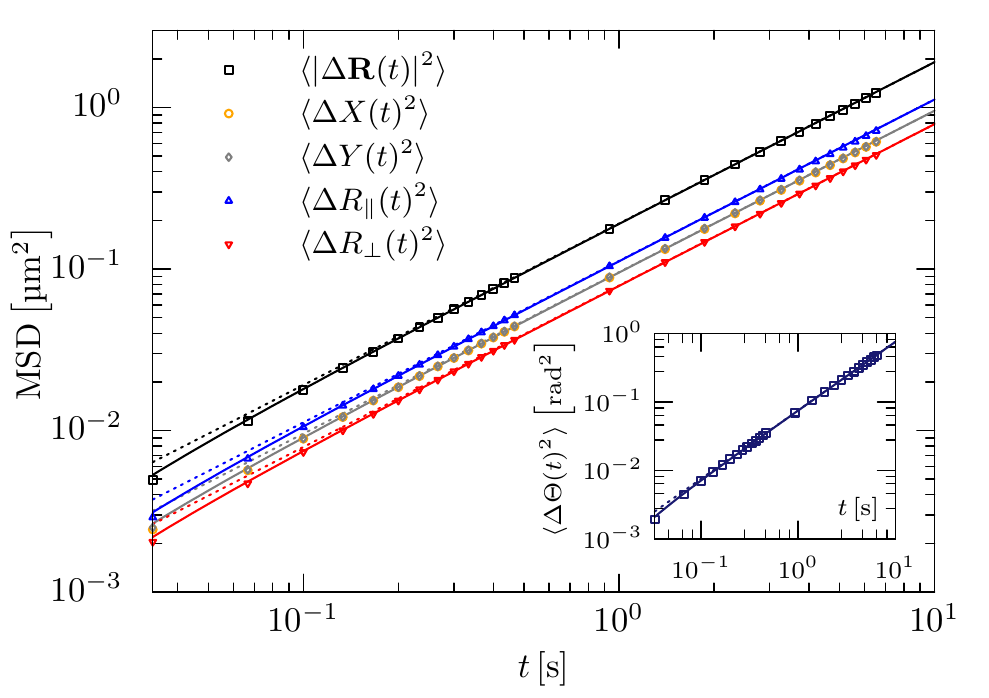}
\caption{Mean-square angular displacement (MSAD) $\langle \Delta \Theta(t)^2 \rangle$ (inset) and mean-square displacement (MSD) of dimers  in the laboratory frame in the plane $\langle | \Delta \vec{R}(t)|^2\rangle$, 
 along the two cartesian directions $\langle\Delta X(t)^{2}\rangle$ and $\langle\Delta Y(t)^{2}\rangle$ (falling on top of each other) and in the particle frame parallel $\langle\Delta R_{\parallel}(t)^2\rangle$ and perpendicular $\langle\Delta R_{\perp}(t)^2\rangle$ to the principal axis. 
The open symbols correspond to experimental data with only subsets of the data points shown,  lines represent  theoretical predictions with the transport coefficients determined by fits to $\langle\Delta R_{\parallel}(t)^2\rangle, \langle\Delta R_{\perp}(t)^2\rangle$ and $\langle \Delta \Theta(t)^2 \rangle$ 
whereas $\langle | \Delta \vec{R}(t)|^2\rangle$, $\langle\Delta X(t)^{2}\rangle$ and $\langle\Delta Y(t)^{2}\rangle$  are calculated without any free parameters based on the determined transport coefficients (see main text).
The theoretical calculations do not (dashed lines) or do (solid lines) take into account dynamic errors.
}
\label{fig:plotMSD1}
\end{figure}

\subsection{Mean-square (angular) displacements and transport coefficients}

First, we calculate the mean-square angular displacement (MSAD) $\langle\Delta\Theta(t)^{2}\rangle$.
In this calculation, as in the calculations of the other correlation functions, we profit from an efficient blocking scheme~\cite{Colberg:CompPhysComm_182:2011}.
Since the angles vary from frame to frame only slightly (approximately $0.04\,\text{rad}= 2.29^{\circ}$),  the angle $\Theta(t)$ can be determined without the need to fold back to the interval $[0,2\pi)$. 
 The MSAD shows the theoretically expected  diffusive behavior $\langle \Delta \Theta(t)^2 \rangle = 2 D_{\text{rot}} t$ (inset Fig.~\ref{fig:plotMSD1}) taking into account dynamic errors (see Appendix~\ref{app:Exposuretime} for details).
This allows us to extract the  rotational self-diffusion coefficient  $D_{\text{rot}}=(3.77\pm 0.02) \times 10^{-2}\, \text{s}^{-1}$.
The characteristic rotational time hence is $\tau_\text{rot} = 26.5\,\text{s}$.

Next, we compute the mean-square displacements (MSD) for the translational motion in the particle frame. By rotating each frame to the current orientation, the translational increments to the next frame can be projected parallel and perpendicular to the current orientation (see Appendix ~\ref{app:LabPartframe}). 
 A linear increase $\langle \Delta R_\parallel(t)^2 \rangle = 2 D_\parallel t$ and $\langle \Delta R_\perp(t)^2 \rangle= 2 D_\perp t$ is expected and observed (Fig.~\ref{fig:plotMSD1}), where the corresponding self-diffusion coefficients were determined as  $D_{\parallel}=(5.61\pm 0.02)\times 10^{-2}\,\text{\textmu{}m}^{2}\text{s}^{-1}$ and $D_{\perp}=(3.95\pm 0.01)\times 10^{-2}\,\text{\textmu{}m}^{2}\text{s}^{-1}$.
The three transport coefficients $D_{\parallel}, D_{\perp}$ and $D_{\text{rot}}$ are determined by a linear least-square fit and the cited uncertainties are the standard errors of the mean over all trajectories. The fits include a correction of the theoretical predictions for dynamic errors~\cite{Savin:BiophysJ_88:2005, Savin:PRE_71:2005} (see  Appendix~\ref{app:Exposuretime}). They affect the MS(A)D at short times, which is  visible in Fig.~\ref{fig:plotMSD1}. In addition, in the fits we discard data points at large lag times ($t > 10\,\text{s}$), since they are based on only a small number of realizations and hence are statistically less reliable or are subject to a slight drift.

The mean self-diffusion coefficient can now be calculated as $\bar{D}=(D_{\parallel} + D_{\perp})/2 = (4.78\pm 0.01)\times 10^{-2}\, \text{\textmu}\text{m}^2\text{s}^{-1}$ and the translational anisotropy 
 to $\Delta D=(D_{\parallel} - D_{\perp}) = (1.66\pm 0.03)\times 10^{-2}\,\text{\textmu}\text{m}^2\text{s}^{-1}$ 
resulting in an anisotropy $\Delta D/\bar{D}=(0.35\pm 0.01)$.
Moreover,  the  MSD in the two cartesian directions of the laboratory frame $\langle \Delta X(t)^2 \rangle=  \langle \Delta Y(t)^2 \rangle = 2 \bar{D} t$ 
 taking into account dynamic errors,  agree well with the experimental data (Fig.~\ref{fig:plotMSD1}).

\subsection{Crossover between isotropic and anisotropic diffusion}

The MSD for any two cartesian directions in the laboratory frame do not display  features connected to the TR coupling (Fig.~\ref{fig:plotMSD1}). However, by restricting the initial orientation, valuable information can be extracted.
 We set the initial angle as $\Theta(0)= \vartheta_0=0$, i.e.\@ the $x$-direction is always aligned along the initial orientation of the particle. 
Then Eqs.~\eqref{eq:MSDComp1} and \eqref{eq:MSDComp2}
 predict a diffusive growth for the two components $\langle\Delta X(t)^{2}\rangle_{0}\simeq 2D_{\parallel}t$ and $\langle\Delta Y(t)^{2}\rangle_{0}\simeq 2D_{\perp}t$ in the time regime $t\ll\tau_{\text{rot}}$. 
For long times $t\gg \tau_{\text{rot}}$ these two restricted MSD approach the average diffusion $\langle \Delta X(t)^2 \rangle_0 \simeq \langle \Delta Y(t)^2 \rangle_0 \simeq 2 \bar{D}t$, where the detailed shape of the crossover is encoded in $\tau_4(t)$, Eq.~\eqref{eq:tau_n}.
To highlight the TR coupling we introduce the  time-dependent self-diffusion coefficients 
\begin{subequations}
\begin{align}
D_{xx}(t)&=\frac{1}{2}\frac{\textrm{d}}{\textrm{d}t}\langle\Delta X(t)^{2}\rangle_{0}, \\
D_{yy}(t)&=\frac{1}{2}\frac{\textrm{d}}{\textrm{d}t}\langle\Delta Y(t)^{2}\rangle_{0}.
\end{align}
\end{subequations}
The quantities $D_{xx}(t)$ and $D_{yy}(t)$ are obtained by finite differences after interpolating the MSD on the sparse grid with a linear spline.

Equations~\eqref{eq:MSDComp1} and \eqref{eq:MSDComp2} encode a symmetric splitting for properly shifted and scaled time-dependent self-diffusion coefficients
\begin{align}\label{eq:Norm_Diffusion}
\frac{D_{xx}(t) - \bar{D}}{\Delta D/2 } = - \frac{D_{yy}(t) - \bar{D}}{\Delta D/2 } = \exp(- 4 D_{\text{rot}}t ) .
\end{align}
The transition from the symmetric split $\pm 1$ to the common $0$ occurs on the time scale  $\tau_{\text{rot}}$.

\begin{figure}[htbp]
\centering
\includegraphics[width=0.5\textwidth]{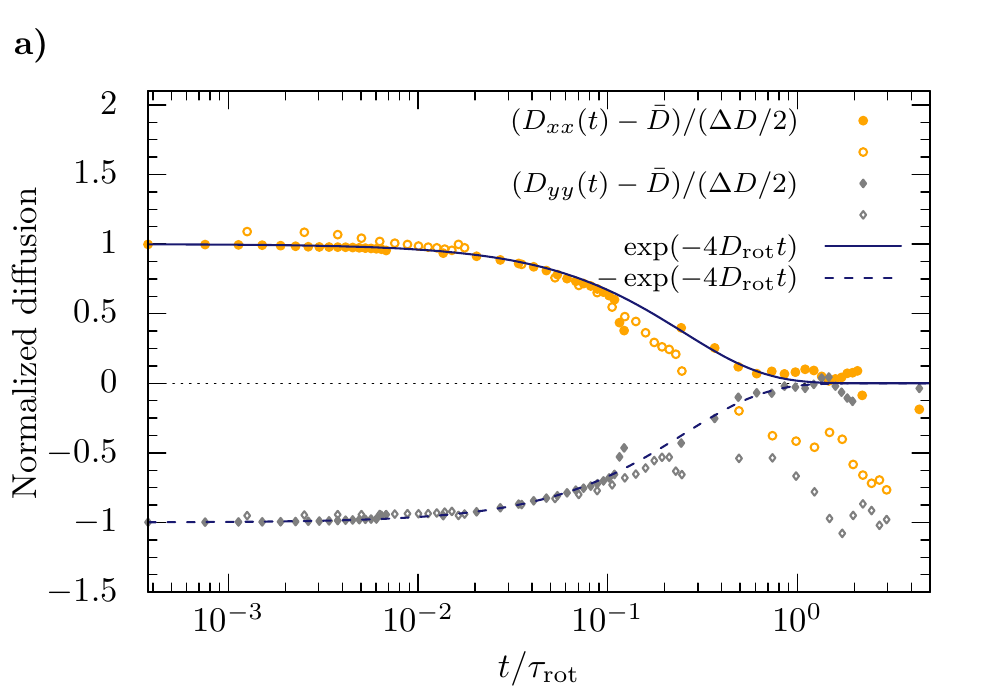}
\includegraphics[width=0.5\textwidth]{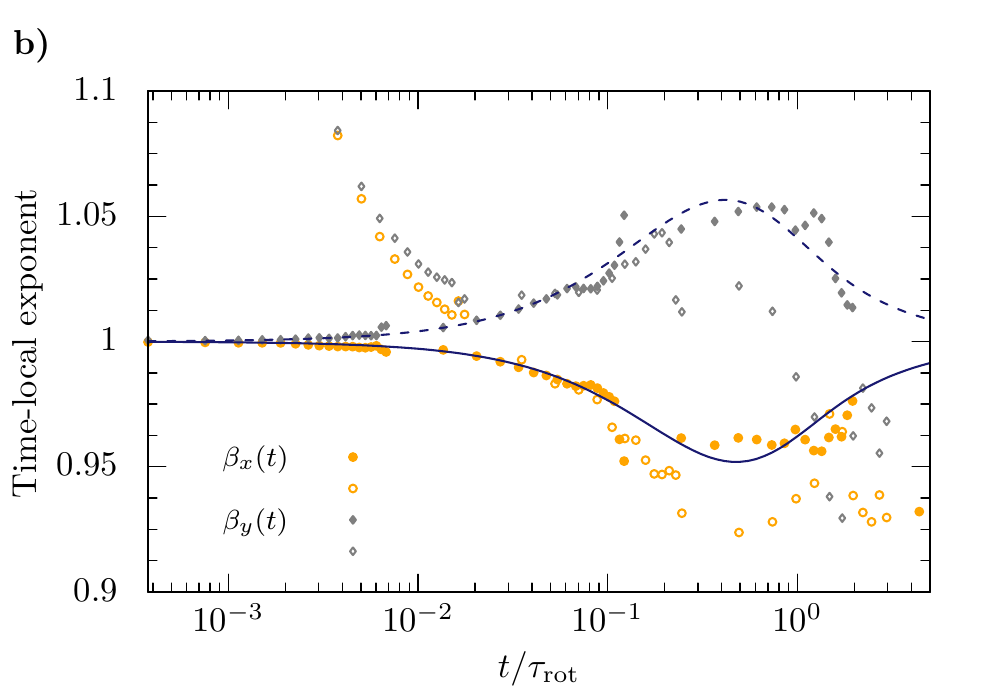}
\caption{\textbf{a)} Normalized time-dependent self-diffusion coefficients $(D_{xx}(t)-\bar{D})/(\Delta D/2)$ (circles) and $(D_{yy}(t)-\bar{D})/(\Delta D/2)$ (diamonds) and \textbf{b)} time-local exponents $\beta_{x}(t)$ (circles) and $\beta_{y}(t)$ (diamonds) of dimers in the laboratory frame. The  $x$-axis was chosen to  align with the initial dimer orientation $(\vartheta_{0}=0)$. Shown are results from experiment (open symbols), stochastic simulations (filled symbols) and theory (lines).}
\label{fig:plotDCoef1}
\end{figure}

The experimental results and the theoretical predictions agree 
up to lag times $t \lesssim  0.1 \tau_{\text{rot}} \approx 2.7\,\text{s}$  beyond which  the data displays significant  statistical fluctuations (Fig.~\ref{fig:plotDCoef1}\textbf{a)}). 
The residual discontinuities in the numerical derivative are due to the blocking
scheme~\cite{Colberg:CompPhysComm_182:2011}.
Our results are consistent with previous studies on   ellipsoidal particles~\cite{Han:Science_314:2006}.
We also compute the time-local scaling exponents via
\begin{subequations}
\begin{align}
\beta_{x}(t)&=\frac{\diff\log\left[\langle\Delta X(t)^2\rangle_{0}\right]}{\diff\log(t)}, \\
\beta_{y}(t)&=\frac{\diff\log\left[\langle\Delta Y(t)^2\rangle_{0}\right]}{\diff\log(t)}.
\end{align}
\end{subequations}
These quantities indicate the transition from initial diffusion $\beta_{x}(t)\simeq\beta_{y}(t)\simeq 1$ for $t\ll\tau_{\text{rot}}$, to small deviations from diffusive behavior at intermediate times $t\approx\tau_{\text{rot}}$, to re-established diffusion at long times $t\gg\tau_{\text{rot}}$. Unfortunately, they turn out to be rather sensitive to dynamic errors (see Fig.~\ref{fig:plotDCoef1}\textbf{b)}), such that the short time behavior of the experimental results is not well resolved and for lag times $t \gtrsim  0.1 \tau_{\text{rot}} \approx 2.7\,\text{s}$, the data again displays significant statistical fluctuations.

To validate our results, we also numerically solve the Langevin equations, Eqs.~\eqref{eq:LangevinAngle} and \eqref{eq:LangevinPos},
 with transport coefficients measured in the experiments as input. We then  calculate the time-dependent self-diffusion coefficients and the time-local exponents with the same algorithm as for the experimental data and find  agreement with the experimental observations and theoretical predictions (Fig.~\ref{fig:plotDCoef1}).

\subsection{Generalized self-intermediate scattering functions}

\begin{figure}[htbp]
\centering
\includegraphics[width=0.5\textwidth]{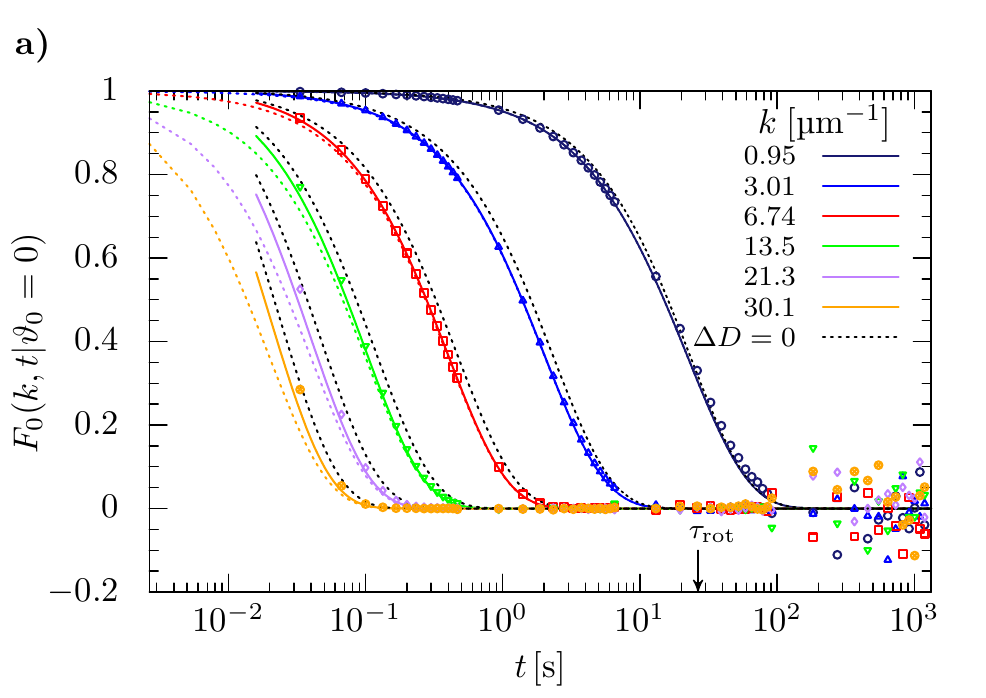}
\includegraphics[width=0.5\textwidth]{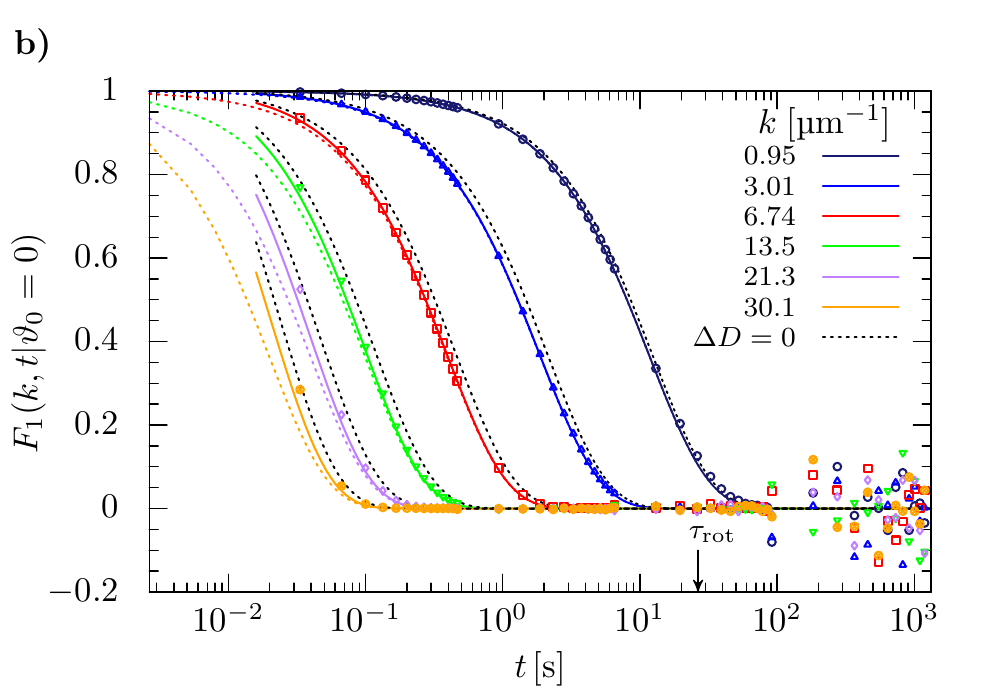}
\caption{Restricted self-intermediate scattering functions \textbf{a)} $F_{0}(k,t\vert\vartheta_{0}{=}0)$ and \textbf{b)} $F_{1}(k,t\vert\vartheta_{0}{=}0)$ as a function of lag time $t$
for various wavenumbers $k$. Experimental data (symbols) and theoretical predictions with (solid lines) and without (dashed lines) correction for dynamic errors, as well as theoretical predictions for isotropic diffusion ($D_{\parallel}=D_{\perp}$) 
and initial orientation $\vartheta_{0}=0$, Eq.~\eqref{eq:ISFiso1}, with correction for dynamic errors (black dashed lines).
}
\label{fig:plotISF1}
\end{figure}

More general spatio-temporal information is encoded in the restricted ISF $F_{\nu}(k,t|\vartheta_{0}{=}0) = \langle \exp( -\i k \Delta X(t) -\i \nu \Theta(t)  ) \rangle_0$,
which is evaluated for  
$\nu=0$ and $1$.
We consider lag times $t$ up to several minutes, i.e.\@ much longer than the rotational diffusion time $\tau_{\text{rot}}= 26.5\,\text{s}$, 
and  over a  wide range of wavenumbers  $0.95\,\text{\textmu{}m}^{-1} \leq k \leq 
30.1\,\text{\textmu{}m}^{-1}$ allowing us to resolve  dynamic processes on length scales much smaller than and comparable to the geometric size of the dimers, approximately $4\,\text{\textmu{}m}$ along the major  axis.

The restricted ISF fall off monotonically for all wavenumbers and resemble an exponential decay (Fig.~\ref{fig:plotISF1}). In particular, for small wavenumbers they are well approximated by $\exp\left[-(\mu^{2} D_{\text{rot}}+k^{2}\bar{D})t\right]$, which represents the description without  TR coupling. This suggests that the spatial and angular degrees of freedom are independent on large length scales as for  isotropic particles with $\Delta D=0$, Eq.~\eqref{eq:ISFiso1}. For small wavenumbers this behavior is expected, since 
the deformation parameter is small $q\lesssim 1$ ($k \lesssim 3.01\, \text{\textmu{}m}^{-1}$) and the Mathieu functions $\text{ce}_{n}(\theta, q)$ become close to pure cosines such that the influence of other modes is negligible.  
In contrast, increasing  the wavenumber small deviations  from a pure exponential relaxation become visible. Although they only amount to a few percent, they are significantly beyond the noise level. The ISF are correctly described by the theoretical predictions if corrections for dynamic errors are taken into account, which are particularly important at very short times~\cite{Savin:BiophysJ_88:2005, Savin:PRE_71:2005}  (Appendix~\ref{app:Exposuretime}).   
The experimental observation of TR coupling and the successful theoretical description in terms of the restricted ISF is one of the main results of this work.

\begin{figure}[htbp]
\centering
\includegraphics[width=0.5\textwidth]{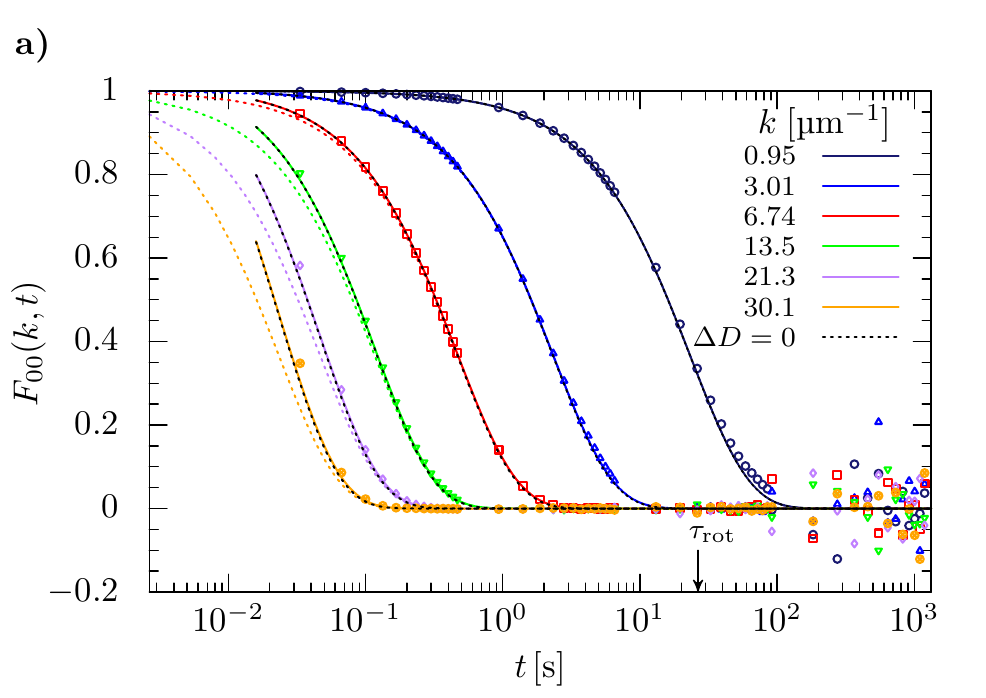}
\includegraphics[width=0.5\textwidth]{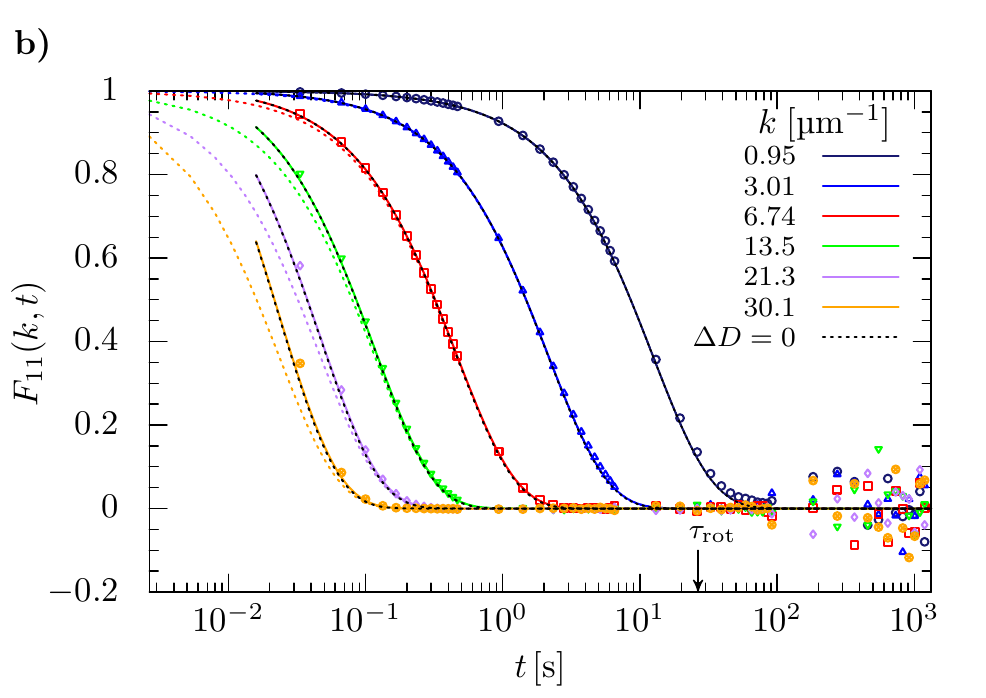}
\caption{Diagonal elements of the generalized self-intermediate scattering functions \textbf{a)} $F_{00}(k,t)$ and \textbf{b)} $F_{11}(k,t)$  as a function of lag time $t$
for various wavenumbers $k$. Experimental data (symbols) and theoretical predictions with (solid lines) and without (dashed lines) correction for dynamic errors, as well as theoretical predictions for isotropic diffusion ($D_{\parallel}=D_{\perp}$) 
and initial orientation $\vartheta_{0}=0$, Eq.~\eqref{eq:ISFiso1}, with correction for dynamic errors (black dashed lines). 
}
\label{fig:plotISF2}
\end{figure}

We also determined diagonal elements of the generalized ISF, Eq.~\eqref{eq:ISF2}, namely $F_{00}(k,t) = F(k,t) = \langle\exp(-\i k \Delta X(t) )\rangle$, which corresponds to the conventional self-intermediate scattering function, Eq.~\eqref{eq:convISF},  
and $F_{11}(k,t) = \langle\exp(-\i  k \Delta X (t)-\i \Delta\Theta(t) )\rangle$.
The data quantitatively agree with the theoretical predictions (Fig.~\ref{fig:plotISF2}).
However, the decays of $F_{00}(k,t)$ and $F_{11}(k,t)$ do not significantly deviate from an exponential decay even at small wavenumbers $k$. Hence the effects of the TR coupling cannot be detected given the typical noise of the data (Appendix~\ref{app:Correlation_functions}).
This is a marked difference to the restricted self-intermediate scattering functions $F_{0}(k,t\vert\vartheta_{0}{=}0)$ and $F_{1}(k,t\vert\vartheta_{0}{=}0)$.

\begin{figure}[htbp]
\centering
\includegraphics[width=0.5\textwidth]{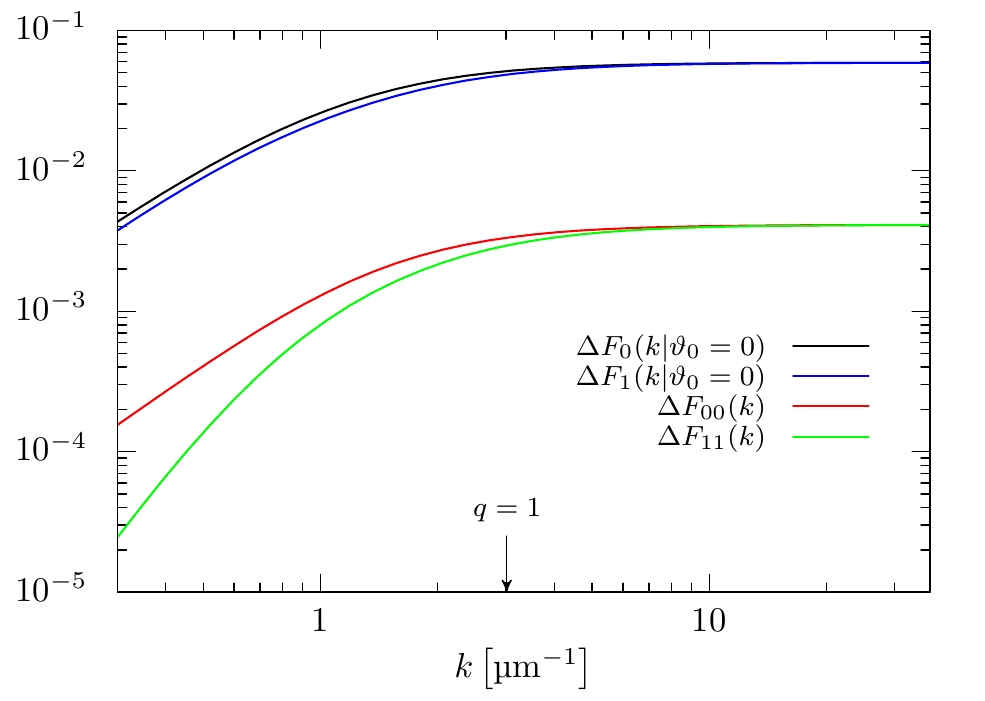}
\caption{Maximum absolute difference between the predictions of the isotropic diffusion model, i.e.\@ an exponential decay, and the anisotropic diffusion theory for the restricted ISF $\Delta F_{0}(k | \vartheta_{0}{=}0), \Delta F_{1}(k | \vartheta_{0}{=}0)$ and for the diagonal elements of the generalized ISF $\Delta F_{00}(k), \Delta F_{11}(k)$ as a function of wavenumber $k$. The wavenumber $k= 3.014 \,\text{\textmu{}m}^{-1}$ corresponds to the deformation parameter $q=1$. 
}
\label{fig:plotDeltaISF}
\end{figure}

It is instructive to elaborate  where  deviations from a pure exponential relaxation 
are expected and in which regimes they become experimentally accessible. 
Therefore we calculate the maximum absolute difference between the theoretical predictions  for $F_{\nu}(k,t|\vartheta_{0}{=}0)$ and the isotropic diffusion model, (Eq.~\eqref{eq:ISFiso1},
\begin{align}\label{eq:DeltaF}
\lefteqn{\Delta F_{\nu}(k|\vartheta_{0}{=}0)= \nonumber } \\
&=\underset{t}{\text{max}}\left[ | F_{\nu}(k,t|\vartheta_{0}{=}0)-F_{\nu}^{\text{Iso}}(k,t\vert\vartheta_{0}{=}0)|\right]. 
\end{align}  
The maximum difference between a diagonal element of the generalized ISF $F_{00}(k,t)$ or $F_{11}(k,t)$ and an exponential decay  is defined correspondingly.
These quantities  measure the quality of approximating  an ISF by a single exponential as in Eqs.~\eqref{eq:ISFiso1} and \eqref{eq:ISFiso2}.  
The restricted ISF $F_{\nu}(k,t|\vartheta_{0}{=}0)$ display deviations of up to $6\,\%$, the diagonal elements of the generalized ISF $F_{00}(k,t)$ and $F_{11}(k,t)$ display deviations that remain well below $1\,\%$ (Fig.~\ref{fig:plotDeltaISF}).
The largest deviations occur around the characteristic time $\tau_\text{c}$ when the ISF decayed to about $1/\e$ (Figs.~\ref{fig:plotISF1} and \ref{fig:plotISF2}).
The deviations become  more important as the inverse length scale quantified by the wavenumber $k$ increases and saturate beyond $k \approx 3.01\, \text{\textmu{}m}^{-1}$ corresponding to a deformation 
parameter of $q = 1$.
These deviations can be compared to typical uncertainties in the experiments (Appendix~\ref{app:Correlation_functions}).
 For the restricted and generalized ISF, the uncertainty $\epsilon \approx 0.1\,\%$ with only minor differences between $F_0(k,t|\vartheta_{0}{=}0)$, $F_1(k,t|\vartheta_{0}{=}0)$, $F_{00}(k,t)$ and $F_{11}(k,t)$ as well as a negligible dependence on $k$.
This is consistent with the possibility to observe TR coupling in the restricted ISF (Fig.~\ref{fig:plotISF1}) and the difficulty to detect TR coupling in the diagonal elements of the generalized ISF (Fig.~\ref{fig:plotISF2}).
Furthermore, the small experimental uncertainties are reflected in only very limited random fluctuations in the data, which only increase towards large $t$ due to the very small number of statistically independent pairs that are available to calculate the ISF.

\begin{figure}[htbp]
\centering
\includegraphics[width=0.5\textwidth]{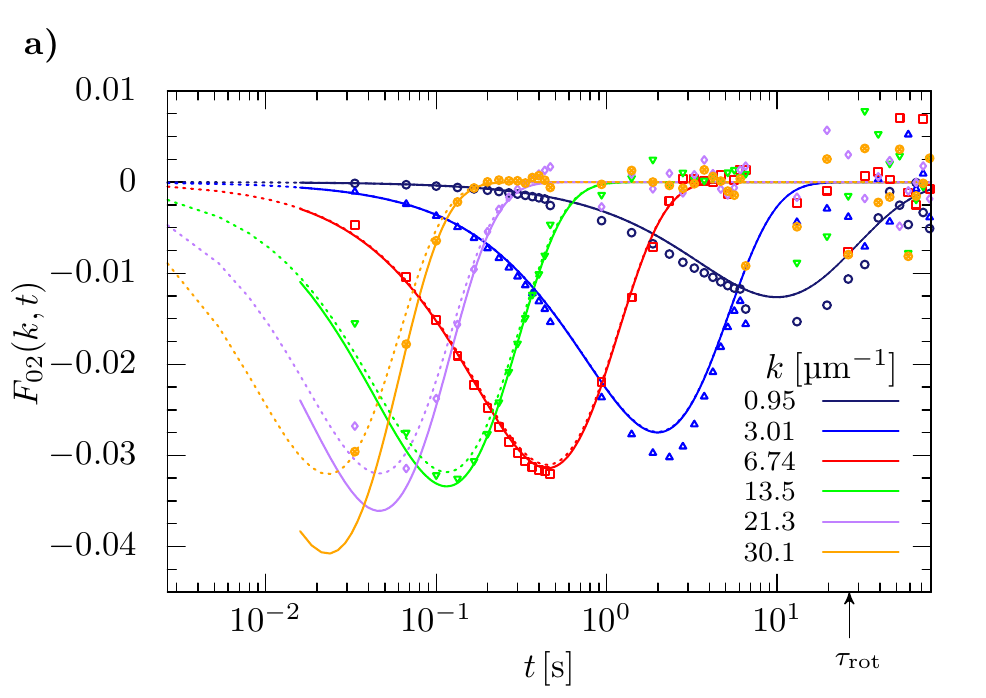}
\includegraphics[width=0.5\textwidth]{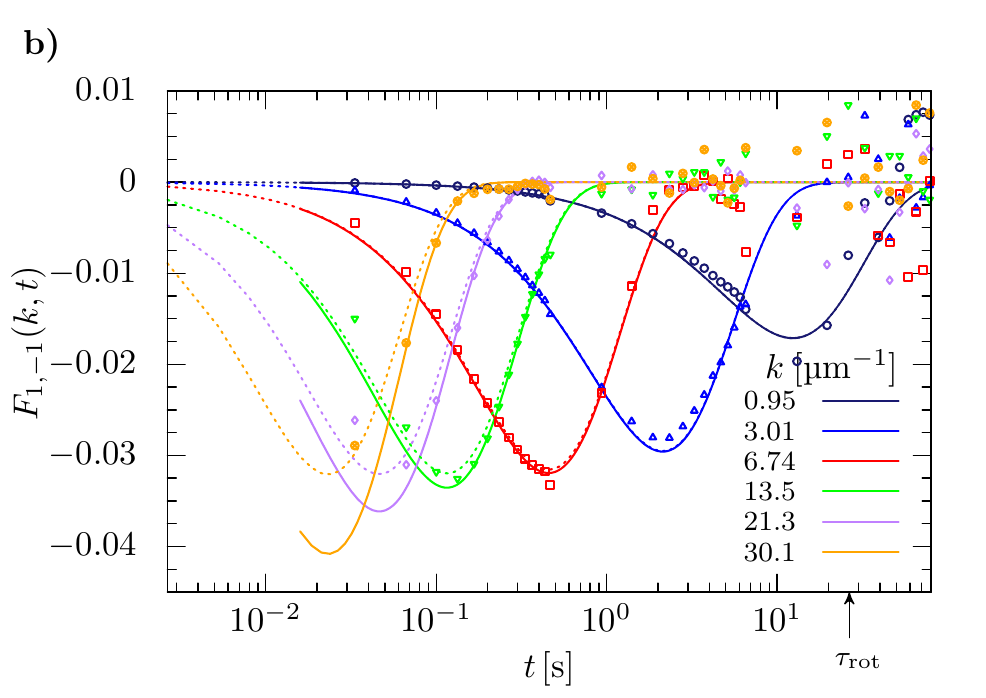}
\caption{Off-diagonal elements of the generalized self-intermediate scattering functions \textbf{a)} $F_{02}(k,t)$ and \textbf{b)} $F_{1,-1}(k,t)$ as a function of lag time $t$ for various wavenumbers $k$. Experimental data (symbols) and theoretical predictions with (solid lines) and without (dashed lines) correction 
for dynamic errors. 
}
\label{fig:plotISF3}
\end{figure}

The off-diagonal elements of the generalized ISF $F_{\mu\nu}(k,t) =\langle\exp(-\i\left(k \Delta X(t)-\mu\Theta(0)+\nu\Theta(t)\right))\rangle$, Eq.~\eqref{eq:ISF2}, vanish
 without TR coupling.  Any signal beyond noise thus is a clear fingerprint of the TR coupling. By the selection rule $(\mu-\nu) \in 2 \mathbb{Z}$, the nontrivial off-diagonal elements with smallest mode index are $F_{02}(k,t)$ and $F_{1,-1}(k,t)$. The theoretical predictions all start at zero and exhibit a prominent
 minimum (Fig.~\ref{fig:plotISF3}). Its position moves to shorter times $t$ and its magnitude increases  as the wavenumber $k$  is increased.  The magnitude of the minimum remains small for all wavenumbers, only a
 few percent. Nevertheless, it is significantly larger than the expected experimental uncertainty, about $0.1\,\%$, and the observed noise in the data. The experimental data quantitatively agree with the theoretical results with relative deviations remaining below $10\,\%$.

We have also computed theoretical predictions for the off-diagonal elements with next-larger indices, $F_{13}(k,t), F_{2,-2}(k,t)$ and $F_{04}(k,t)$ (not shown). They all display  similar behavior as the off-diagonal elements of the generalized ISF with smallest mode index. In particular, $F_{13}(k,t)$  displays a minimum of the same magnitude for all wavenumbers. Interestingly, this changes for $F_{2,-2}(k,t)$ and $F_{04}(k,t)$, where we find maxima with a much smaller amplitude of only a few tenths of a percent. This is no longer resolvable in the experimental data given the typical experimental uncertainties of a tenth of a percent. In general, whether a maximum or minimum occurs, depends on the choice of the mode indices. However, the decrease of their amplitudes with increasing mode indices seems to be a common feature.

\section{Summary and Conclusion}\label{Sec:Summary}

The dynamics of colloidal dimers in 2D have been investigated by single-particle tracking and compared to theoretical predictions for anisotropic diffusion. We have elaborated the theory for measurable quantities that are sensitive to  translational-rotational (TR) coupling. 
In particular,  we investigate the dynamics at different length scales. This spatio-temporal behavior is probed as a function of the wavenumber using the (generalized) self-intermediate scattering functions.
 Their long-wavelength properties contain as special case the low-order moments previously studied by Han et al.~\cite{Han:Science_314:2006}.

The mean-square displacements (MSD) for the two cartesian directions in the laboratory frame, $\langle \Delta X(t)^2 \rangle$ and $\langle \Delta Y(t)^2 \rangle$, do not show any feature connected to TR coupling. However,
 the mean-square displacements restricted to the initial orientation $\vartheta_0 = 0$, $\langle \Delta X(t)^2 \rangle_0$ 
and $\langle \Delta Y(t)^2 \rangle_0$, as well as  the corresponding time-dependent self-diffusion coefficients, $D_{xx}(t)$ and $D_{yy}(t)$, show clearly visible effects. In addition,  the restricted self-intermediate scattering functions show deviations from a single-exponential decay. Despite their small size, only a few percent, they are accessible in our experiments which yield the restricted ISF with an uncertainty of about $0.1\,\%$. 
Furthermore, we show that the off-diagonal elements of the generalized self-intermediate scattering function exhibit a distinct fingerprint of the anisotropic diffusion and thus the TR coupling.
Although their magnitude  also amounts to only a few percent, it is beyond the experimental uncertainty  and we have  observed it in our experiments beyond the noise level. 
In contrast, the conventional self-intermediate scattering function $F(k,t) = F_{00}(k,t)$ or, more general, the diagonal elements of the generalized self-intermediate scattering function $F_{\nu\nu}(k,t)$ differ from an exponential decay only very slightly, by a few tenths of a percent.
This is about the noise level in our experiments indicating that it is very difficult to obtain experimental data with a low enough noise level that the effects of TR coupling become detectable using particle tracking.
The quantitative agreement between theoretical predictions and experimental results provides confidence in our calculations as well as the quality of the experimental data.

Scattering experiments typically yield superior statistics compared to tracking experiments.
However, they neither provide access to the restricted mean-square displacements nor to the restricted self-intermediate scattering functions.
To observe TR coupling in scattering experiments, thus the noise level needs to be low enough to detect the much smaller differences between the anistropic and isotropic diffusion in the conventional self-intermediate scattering functions.
Hence this represents a very significant experimental challenge.
In this context it is interesting to note that the data analysis of depolarized-light-scattering experiments is typically based on a decoupling approximation. It is assumed that the decays caused by the translational and rotational motions are independent. This allows one to  determine the translational and rotational dynamics due to their different dependences on the wavenumber $k$ \cite{Brogioli:OE_17:2009, Escobedo:JCP_17:2015}.
This assumption appears justified due to the small effect of TR coupling.

In general, the deviations from a pure exponential decay remain small for the dimers investigated.
Increasing the relative anisotropy by increasing the elongation of the particles may  give rise to more pronounced and better quantifiable effects.
Moreover, other anisotropic particles, such as molecules~\cite{Berger:JPCB_123:2019, Cichocki:JFM_878:2019}, proteins~\cite{Smith:JMB_236:1994,Yamamoto:PhysRevLett_126:2021},  elliptical~\cite{Han:PRE_80:2009}, or `boomerang' particles~\cite{Chakrabarty:Langmuir_30:2014, Chakrabarty:SM_12:2016, Koens:SM_13:2017} are expected to  display similar TR coupling possibly with only a slightly larger effect on the ISF. Note that for particles belonging to the symmetry point group $D_{3}$ (dihedral group), such as e.g.\@ trimers, the translational anisotropy  and effects due to TR coupling vanish \cite{Chakrabarty:Langmuir_30:2014}.  In any case, measuring the generalized self-intermediate scattering functions should provide quantitative information on their transport behavior.  Our theory could be extended to investigate their TR coupling, also in the presence of additional external forces.

 We anticipate much more drastic effects for active agents,
 such as catalytic L-shaped Janus particles~\cite{Kuemmel:PRL_110:2013,tenHagen:NatComm_5:2014}, or anisotropic particles in active, e.g. bacterial, suspensions \cite{Peng:PRL_116:2016}. Extending the theoretical predictions for the generalized self-intermediate scattering functions to an anisotropic active Brownian particle~\cite{Kurzthaler:SM_13:2017} should provide valuable insights into the TR coupling of active agents  and, e.g.,  allow elucidating  their gravitactic behavior~\cite{tenHagen:NatComm_5:2014}.

\section*{acknowledgements}
This work has been supported  by the D-A-CH project `Structure and Dynamics of Liquids in Confinement' funded by the Austrian Science Fund (FWF): I 2887 (T.F.~and D.B.M) and the Deutsche Forschungsgemeinschaft (DFG):  EG~269/7-1 (S.U.E.~and M.A.E.S.).
C.K.~acknowledges support from the Austrian Science Fund (FWF) via the Erwin Schr{\"o}dinger fellowship (Grant No.~J4321-N27).

\appendix

\section{Numerical solution of the eigenvalue problem for the Fourier coefficients }
\label{app:numMathieu}

The Mathieu functions are essentially deformed cosine and sine functions with Fourier representation~\cite{Nist:Handbook_of_Mathematical_Functions:2010, Nist:website_Mathieu:2010}
\begin{subequations}
\begin{align}
\text{ce}_{2n}(\vartheta, q)&=\sum_{m=0}^{\infty}A_{2m}^{2n}(q)\cos(2m\vartheta),\label{eq:Fourier1} \\
\text{ce}_{2n+1}(\vartheta, q)&=\sum_{m=0}^{\infty}A_{2m+1}^{2n+1}(q)\cos((2m{+}1)\vartheta), \\
\text{se}_{2n+2}(\vartheta, q)&=\sum_{m=0}^{\infty}B_{2m+2}^{2n+2}(q)\sin((2m{+}2)\vartheta), \\
\text{se}_{2n+1}(\vartheta, q)&=\sum_{m=0}^{\infty}B_{2m+1}^{2n+1}(q)\sin((2m{+}1)\vartheta),\label{eq:Fourier4}
\end{align}
\end{subequations}
where the coefficients and eigenvalues are evaluated numerically by inserting the Fourier series  into the standard Mathieu equation. For even mode indices the following eigenvalue 
problems have to be solved~\cite{Ziener:JCompAppMath_236:2012, Chador:RevMexFis_48:2002}
\begin{subequations}
\begin{align}
{\sf M}_{1}^{2n}\cdot\mathbf{A}^{2n}&=a_{2n}\mathbf{A}^{2n}, \\
{\sf M}_{2}^{2n+2}\cdot\mathbf{B}^{2n+2}&=b_{2n+2}\mathbf{B}^{2n+2}
\end{align}
\end{subequations}
with eigenvectors
\begin{subequations}
\begin{align}
\mathbf{A}^{2n}(q)&=\left(\sqrt{2}A_{0}^{2n}, A_{2}^{2n}, \cdots, A_{2M}^{2n}\right)^{\text{T}}, \\
\mathbf{B}^{2n+2}(q)&=\left(B_{2}^{2n+2}, B_{4}^{2n+2}, \cdots, B_{2M+2}^{2n+2}\right)^{\text{T}}
\end{align}
\end{subequations}
and corresponding band matrices
\begin{subequations}
\begin{align}
{\sf M}_{1}^{2n}(q)&=\begin{pmatrix}
                0            & \sqrt{2}q & 0    &           & 0  \\
               \sqrt{2}q & 4            & q     &\ddots    & \\
               0             & q            & 16  & \ddots &  0 \\
                           & \ddots             &   \ddots &\ddots  &  q   \\
               0             &               &     0& q          & 4M^2\\
               \end{pmatrix}, \\         
{\sf M}_{2}^{2n+2}(q)&=\begin{pmatrix}
                4           & q & 0    &            & 0  \\
               q & 16            & q    &\ddots  & \\
               0             & q            & 36 & \ddots &  0 \\
                              & \ddots              &   \ddots        &\ddots  &  q   \\
               0             &                 &     0     & q & (2M+2)^2\\
               \end{pmatrix}.     
\end{align}
\end{subequations}
For odd mode indices we deal with \cite{Chador:RevMexFis_48:2002}
\begin{subequations}
\begin{align}
{\sf M}^{2n+1}_{1}\cdot\mathbf{A}^{2n+1}&=a_{2n+1}\mathbf{A}^{2n+1}, \\
{\sf M}^{2n+1}_{2}\cdot\mathbf{B}^{2n+1}&=b_{2n+1}\mathbf{B}^{2n+1},
\end{align}
\end{subequations}
where the eigenvectors read
\begin{subequations}
\begin{align}
\mathbf{A}^{2n+1}(q)&=\left(A_{1}^{2n+1}, A_{3}^{2n+1}, \cdots, A_{2M+1}^{2n+1}\right)^{\text{T}}, \\
\mathbf{B}^{2n+1}(q)&=\left(B_{1}^{2n+1}, B_{3}^{2n+1}, \cdots, B_{2M+1}^{2n+1}\right)^{\text{T}}
\end{align}
\end{subequations}
and the band matrices have the following form
\begin{subequations}
\begin{align}
{\sf M}^{2n+1}_{1}(q)&=\begin{pmatrix}
                1+q           & q & 0    &            & 0  \\
               q             & 9            & q    &\ddots  & \\
               0             & q            & 25  & \ddots &  0 \\
                              & \ddots              &   \ddots        &\ddots  &  q   \\
               0             &                 &     0     & q & (2M+1)^2\\
               \end{pmatrix},   \\          
{\sf M}^{2n+1}_{2}(q)&=\begin{pmatrix}
                1-q           & q & 0    &            & 0  \\
               q             & 9            & q    &\ddots  & \\
               0             & q            & 25  & \ddots &  0 \\
                              & \ddots              &   \ddots        &\ddots  &  q   \\
               0             &                 &     0     & q & (2M+1)^2\\
               \end{pmatrix}.         
\end{align}
\end{subequations}
The eigenvalue problems are truncated at an appropriate dimension $M$ so that the conditions $F_{\mu}(\mathbf{k},t{=}0\vert\vartheta_{0}{=}0)=1$ and $F_{\mu\nu}(k,t{=}0)=\delta_{\mu\nu}$ are fulfilled to reasonable accuracy. Thereby, we exploit the orthonormality relations for the Fourier coefficients \cite{Ziener:JCompAppMath_236:2012, Nist:Handbook_of_Mathematical_Functions:2010, Nist:website_Mathieu:2010}
\begin{subequations}
\begin{align}
&\sum_{n=0}^{\infty}A_{2p}^{2n}A_{2r}^{2n}=\delta_{pr}-\frac{1}{2}\delta_{0r}\delta_{0p}, \quad p,r\in\mathbb{N}_{0}, \label{equ:Ortho1}\\
&\sum_{n=0}^{\infty}A_{2p+1}^{2n+1}A_{2r+1}^{2n+1}=\delta_{pr}, \\
&\sum_{n=0}^{\infty}B_{2p}^{2n+2}B_{2r}^{2n+2}=\delta_{pr}, \\
&\sum_{n=0}^{\infty}B_{2p+1}^{2n+1}B_{2r+1}^{2n+1}=\delta_{pr}, \quad p,r\in\mathbb{N}. \label{equ:Ortho4}
\end{align}
\end{subequations}

\section{Artefacts due to dynamic errors}\label{app:Exposuretime}

During the tracking procedure a finite exposure time  of the camera leads to a motion blur of the microscope image, such that the average rather than the instantaneous position of the particle is measured. 
Furthermore, a possible jitter in the frame and digitalization rate leads to a `blur' of the moments and correlation functions.
The effects of these dynamic errors are particularly pronounced at small lag times. We follow Savin and Doyle, who have proposed a correction for the theoretical MSD~\cite{Savin:BiophysJ_88:2005, Savin:PRE_71:2005} 
\begin{equation}\label{eq:CorMSD}
\langle\vert\Delta\vec{R}(t)\vert^{2}\rangle=4\bar{D}\left(t-T_{\text{e}}/3\right),
\end{equation}
and similarly for the theoretical MSAD
\begin{equation}
\langle\Delta\Theta(t)^{2}\rangle=2D_{\text{rot}}\left(t-T_{\text{e}}/3\right),
\end{equation}
for times $t \gtrsim T_\text{e}$ with an effective exposure time $T_\text{e}$ that accounts for both effects.
We use $T_{\text{e}}=(1/60)\,\text{s}$, i.e.\@ half the time between two 
images $\Delta t = (1/30)\,\text{s}$, when comparing the theoretical predictions to the experimentally obtained mean-square and 
mean-square-angular displacements (Fig. ~\ref{fig:plotMSD1}).

The correction  has only recently been extended to the ISF (see Supplement of Ref.~\cite{Kurzthaler:PRL_121:2018}).  Provided there is a large separation of timescales $T_{\text{e}}\ll\tau_{\text{rot}}$, the theoretical results for the generalized and restricted ISF are well approximated in the presence of rotational decorrelation by
\begin{subequations}
\begin{align}
F_{\mu}^{\text{cor}}(\vec{k},t\vert\vartheta_{0})&\simeq \e^{\bar{D}k^{2}T_{\text{e}}/3}F_{\mu}(\vec{k},t\vert\vartheta_{0}), \label{eq:CorrestISF} \\
F_{\mu\nu}^{\text{cor}}(k,t)&\simeq\e^{\bar{D}k^{2}T_{\text{e}}/3}F_{\mu\nu}(k,t)\label{eq:CorgeneralISF}
\end{align}
\end{subequations}
for $t\gtrsim T_{\text{e}}$ since the decomposition in Fourier-modes does not alter the derivation of the correction factor arising from the pure diffusion case (see Supplement of Ref.~\cite{Kurzthaler:PRL_121:2018}). We thus compare Eqs.~\eqref{eq:CorrestISF} and ~\eqref{eq:CorgeneralISF} to the experimentally determined self-intermediate scattering functions (Figs.~\ref{fig:plotISF1}, \ref{fig:plotISF2} and \ref{fig:plotISF3}).

\section{Laboratory and particle frame of reference}\label{app:LabPartframe}

From the images  one extracts for each dimer the center of hydrodynamic stress (CoH) positions $\vec{R}(t_{n})=(X(t_{n}), Y(t_{n}))^{\text{T}}$ in the laboratory frame and the angles $\Theta(t_{n})$ relative to the $x$-axis at times $t_{n}$.  Hence between two consecutive images the particles displace 
by $\delta\vec{R}(t_{n}):=\vec{R}(t_{n+1})-\vec{R}(t_{n})$, typically
  $0.07\,\text{\textmu{}m}$,  and their orientation changes by $\delta\Theta(t_{n}):=\Theta(t_{n+1})-\Theta(t_{n})$, typically $0.04\,\text{rad}\approx 2.3^{\circ}$.

A rotation of the laboratory frame displacements $\delta\vec{R}(t_{n})$ yields the positional increments in the particle frame  $\delta\check{\vec{R}}(t_{n})=\mathsf{R}(\Theta_{n})\cdot\delta\vec{R}(t_{n})$, where $\mathsf{R}(\Theta_{n})$ denotes the two-dimensional rotation matrix rotating counterclockwise by the orientation at the beginning of each displacement $\Theta_{n}=\Theta(t_{n})$. We then construct CoH positions $\check{\vec{R}}(t_{n})=(R_{\parallel}(t_{n}), R_{\perp}(t_{n}))^{\text{T}}$ by accumulating increments $\check{\vec{R}}(t_{n})=\sum_{i=1}^{n}\delta\check{\vec{R}}(t_{i})$ for $n=1,2,3,\ldots$ with $R_{\parallel}(t)$  the projection and $R_{\perp}(t)$ the rejection along the instantaneous orientation $\vec{U}(t)$ given by the two principle axes  of the particle in two dimensions~\cite{Chakrabarty:PRL_111:2013, Han:PRE_80:2009}.

\section{Calculation of the correlation functions}\label{app:Correlation_functions}

Time correlation functions encode the dynamic properties of a physical system for different timescales. The time autocorrelation function of an observable $A(\tau)$ for a stationary and ergodic random process is expressed as
\begin{align}
C_{AA}(t)=\lim_{T\rightarrow\infty}\frac{1}{T-t}\int_{0}^{T-t}A(\tau+t)A(\tau)^{*}\,\diff \tau
\end{align}
for lag times $t$, where the asterisk denotes the complex conjugate. Its discretized version reads
\begin{equation}\label{equ:discretizedCor}
 C_{AA}(t_m)\approx\frac{1}{M-m}\sum_{i=0}^{M-m-1}A_{m+i}A_{i}^{*} ,
 \end{equation}
 where $t_m = m\Delta t$ is the lag time  and the state $A_{i}:= A(t_i)= A(i\Delta t)$ of the system is obtained from an equidistant time grid $t_i=i\Delta t,\,i\in\{0,\ldots,M-1\}$ with time step $\Delta t$  and a sufficiently long measurement time $T=(M-1)\Delta t$~\cite{Berne:Dynamic_Light_Scattering:2000, Gardiner:Stochastic_Methods:2009, Doi:Theory_of_Polymer_Dynamics:1986}.

For a given number of trajectories $N_{\text{t}}$ with duration $T$ the relevant moments of the corresponding probability distribution can be calculated  taking the time and ensemble averages
\begin{equation}
\langle \left|\Delta\vec{R}(t_m)\right|^{2n}\rangle=\frac{1}{N_{\text{t}}N}
\sum_{j=1}^{N_{\text{t}}}\sum_{i=0}^{N-1}
\left(\vec{R}_{m+i}^{(j)}-\vec{R}_{i}^{(j)}\right)^{2n} , 
\end{equation}
for $n\in\mathbb{N}$ and $N=M-m$. The positions are defined by $\vec{R}_{i}^{(j)}:= \vec{R}^{(j)}(t_i) = (X_{i}^{(j)},Y_{i}^{(j)})^{\text{T}}$ for each trajectory $j\in\{1,\ldots,N_{\text{t}}\}$. Similar expressions hold for the mean-square angular displacement (MSAD) and the mean-square displacement (MSD) of the cartesian components by replacing $\vec{R}(t)$ with the corresponding quantities $\Theta(t)$, $X(t)$ or $Y(t)$, respectively.

For the generalized intermediate scattering function, Eq.~\eqref{eq:ISF2}, with $\vec{k}=k\vec{e}_{x}$ we find
\begin{align}
F_{\mu\nu}(k,t_m)&=\frac{1}{N_{\text{t}}N}\sum_{j=1}^{N_{\text{t}}}\sum_{i=0}^{N-1}\exp\left[-\i k\left(X_{m+i}^{(j)}-X_{i}^{(j)}\right)\right] \nonumber \\
&\times\exp\left[\vphantom{\vec{R}_{m+i}^{(j)}}-\i \left(\nu\Theta^{(j)}_{m+i}-\mu\Theta^{(j)}_{i}\right)\right].
\end{align}
 A moving average along each available trajectory yields the time and ensemble average. Hence the initial angles $\Theta(t_{0})=\vartheta_{0}$ for starting times $t_{0}$ are always different.
Its uncertainty is given by
\begin{align}\label{epsilon_gISF}
\lefteqn{ \epsilon^2(F_{\mu\nu}(k,t_m))  = } \nonumber \\
= &  \sum_{j=1}^{N_\text{t}} \sum_{i=0}^{N-1}\Bigg[   \left| \frac{\partial F_{\mu\nu}}{\partial {X}_{m+i}^{(j)}  } \right|^2 \epsilon^2(X)
+\left| \frac{\partial F_{\mu\nu}}{\partial {X}_{i}^{(j)}  } \right|^2 \epsilon^2(X)
\nonumber \\
&+\left| \frac{\partial F_{\mu\nu}}{\partial {\Theta}_{m+i}^{(j)}  } \right|^2 \epsilon^2(\Theta) 
+\left| \frac{\partial F_{\mu\nu}}{\partial {\Theta}_{i}^{(j)}  } \right|^2\epsilon^2(\Theta) \Bigg]\nonumber \\
=& \left(\frac{1}{N_\text{t} N} \right)^2\sum_{j=1}^{N_\text{t}}  \sum_{i=0}^{N-1} \Big[ 2k^2\epsilon^2(X)+(\nu^2{+}\mu^2)\epsilon^2(\Theta) \Big]\nonumber \\
=& \frac{2}{N_\text{t} N}\left[k^2+\frac{2\,(\nu^2{+}\mu^2)}{(0.9\sigma)^2}\right]\,\epsilon^2(X) ,
\end{align}
where only uncorrelated 
phase factors $\exp\left(-\i k X^{(j)}_{k}\right)$
 contribute to $N$.
We consider
phase factors
 to be uncorrelated if they are separated by a time $\tau^* = \tau^*(k)$, sufficient for the ISF to decay below $1\%$.
Since the ISF for anisotropic and isotropic diffusion are very similar, we base our estimate of $\tau^*$ on the isotropic ISF, Eq.~\eqref{eq:ISFiso2}.
After five times the characteristic time $\tau_\text{c}$, the ISF decayed to below $1\%$, where $\tau^* = 5 \tau_\text{c} = 5 (k^2\bar{D}+\nu^2 D_\text{rot})^{-1}$. 
The uncertainty will be compared with the maximum deviation of the anisotropic and isotropic ISF which occurs around the characteristic time $\tau_\text{c}$ and hence $t_m = \tau_\text{c} \ll T$.
Thus, the number of statistically independent
phase factors can be estimated to be $N = (T-t_m)/\tau^\star \approx (k^2\bar{D}+\nu^2 D_\text{rot}) \, T/5$, where this number is limited by the number of images $T/\Delta t$.

Rotating the trajectory by an angle $\Theta_{i}-\vartheta_{0}$,  each data point can be chosen as an initial point, while ensuring  a fixed orientation $\Theta(t_{0})$. In particular, this can be achieved conveniently during processing the relevant blocks of the blocking scheme~\cite{Colberg:CompPhysComm_182:2011}. 
 The moments of the $X$-component are then computed as
\begin{equation}
\langle \Delta X(t_m)^{2n}\rangle_{\vartheta_{0}}=\frac{1}{N_{\text{t}}N}\sum_{j=1}^{N_{\text{t}}}\sum_{i=0}^{N-1}\left(\tilde{X}_{m+i}^{(j)}-\tilde{X}_{i}^{(j)}\right)^{2n} , 
\end{equation}
 and similarly for the $Y$-component with $n\in\mathbb{N}_{0}$. 
 In addition the restricted ISF, Eq.~\eqref{eq:restricted_ISF}, reads
\begin{align}
F_{\nu}(k,t_m \vert\vartheta_{0})&=\frac{1}{N_{\text{t}}N}\sum_{j=1}^{N_{\text{t}}}\sum_{i=0}^{N-1}\exp\left[-\i k\left(\tilde{X}_{m+i}^{(j)}-\tilde{X}_{i}^{(j)}\right)\right] \nonumber \\
&\times\exp\left[\vphantom{\vec{R}_{m+i}^{(j)}}-\i \nu\left(\Theta^{(j)}_{m+i}-\Theta^{(j)}_{i}+\vartheta_{0}\right)\right],
\end{align}
where $\tilde{\vec{R}}_{k}^{(j)}=(\tilde{X}_{k}^{(j)},\tilde{Y}_{k}^{(j)})^{\text{T}}= \mathsf{R}(\Theta_{i})\cdot\vec{R}_{k}^{(j)}$ 
denote the rotated positions after applying the corresponding rotation matrix
\begin{equation}
\mathsf{R}(\Theta_{i})=\begin{pmatrix}
                                          \cos(\Theta_{i}-\vartheta_{0}) & \sin(\Theta_{i}-\vartheta_{0}) \\
  -\sin(\Theta_{i}-\vartheta_{0})  &  \cos(\Theta_{i}-\vartheta_{0}) \\                                      
                                          \end{pmatrix}.
\end{equation} 
Note that there is a subtle difference compared to the transformation into the particle  frame of reference mentioned in appendix~\ref{app:LabPartframe}. Here, the index $i$  refers to the first data point of the section of the trajectory (i.e.\@ the corresponding blocks in the blocking-scheme) to be rotated.

The uncertainty of the restricted ISF is given by
\begin{align}\label{epsilon_gResISF}
\lefteqn{\epsilon^2(F_{\nu}(k,t_m \vert\vartheta_{0}))  = } \nonumber \\
=&   \sum_{j=1}^{N_\text{t}} \sum_{i=0}^{N-1} \Bigg[ \left| \frac{\partial F_{\nu}}{\partial \tilde{X}_{m+i}^{(j)}  } \right|^2 \epsilon^2(\tilde{X})
+\left| \frac{\partial F_{\nu}}{\partial \tilde{X}_{i}^{(j)}  } \right|^2 \epsilon^2(\tilde{X})
\nonumber \\
&+\left| \frac{\partial F_{\nu}}{\partial {\Theta}_{m+i}^{(j)}  } \right|^2 \epsilon^2(\Theta) +\left| \frac{\partial F_{\nu}}{\partial {\Theta}_{i}^{(j)}  } \right|^2\epsilon^2(\Theta) \nonumber \\
& +\left| \frac{\partial F_{\nu}}{\partial {\vartheta}_0  } \right|^2\epsilon^2(\vartheta_0)  \Bigg] \nonumber \\
=& \left(\frac{1}{N_\text{t} N} \right)^2\sum_{j=1}^{N_\text{t}} \sum_{i=0}^{N-1} \Big[ 2k^2\epsilon^2(X)+ 3 \nu^2\epsilon^2(\Theta) \Big]  \nonumber \\
=& \frac{2}{N_\text{t} N}\left[k^2+\frac{6\nu^2}{(0.9\sigma)^2}\right]\,\epsilon^2(X).
\end{align}
Again, only  uncorrelated
phase factors are considered for $N$.


%

\end{document}